\documentclass[10pt,twoside]{article}
\usepackage{graphicx, subfigure}         
\usepackage{layout}
\usepackage{asa, amscd, natbib2}
\usepackage{amsmath, amsfonts, amssymb}

\usepackage{verbatim}

\newcounter{subfig}

\newcommand{\n}{\noindent}

\begin{document}   

\thispagestyle{empty}

\noindent{\large \bf \em To be published in Technometrics (with Discussion)}

\medskip

\begin{center}

{\bf Julia Brettschneider
\footnotemark[1]\footnotemark[3]${}^{\ast}$,
Fran\c{c}ois Collin\footnotemark[2], Benjamin M.\,Bolstad\footnotemark[2], 
Terence P.\,Speed\footnotemark[2]\footnotemark[4]}

\footnotetext[1]{University of Warwick, Department of Statistics, Coventry, UK;}
\footnotetext[2]{University of California at Berkeley, Department of Statistics, Berkeley, California, USA;}
\footnotetext[3]{Queen's University, Cancer Research Institute Division of Cancer Care $\&$ 
Epidemiology and \vspace{-0.5mm}\\
\phantom{\quad 2}
Department of Community Heath $\&$ Epidemiology, Kingston, Ontario, Canada;}
\footnotetext[4]{Walter and Eliza Hall Institute Bioinformatics Division, Melbourne, 
Australia.\vspace{-0.3mm}\\
\phantom{\quad i}$\ast$Corresponding author: julia.brettschneider@warwick.ac.uk
}


\bigskip

{\Large\bf Quality assessment for}\\
\smallskip
{\Large\bf short oligonucleotide microarray data}\\

\bigskip

\end{center}


\begin{abstract}
\n Quality of microarray gene expression data has emerged as a new research
topic. As in other areas, microarray quality is assessed by comparing 
suitable numerical summaries across microarrays, so that outliers and trends
can be visualized, and poor quality arrays or variable quality sets 
of arrays can be identified. Since each single array comprises tens or 
hundreds of thousands of measurements, the challenge is to find 
numerical summaries which can be used to make accurate quality calls. 
To this end, several new quality measures are introduced based on probe 
level and probeset level information, all obtained as a by-product of the 
low-level analysis algorithms RMA/fitPLM for Affymetrix GeneChips. 
Quality landscapes spatially localize chip or hybridization problems. 
Numerical chip quality measures are derived from the distributions of 
{\it Normalized Unscaled Standard Errors} and of {\it Relative Log Expressions.} 
Quality of chip batches is assessed by {\it Residual Scale Factors.} 
These quality assessment measures are demonstrated on a variety 
of datasets (spike-in experiments, small lab experiments, multi-site 
studies).  They are compared with Affymetrix's individual chip quality report.
\end{abstract}

\n KEYWORDS: quality control, microarrays, Affymetrix chips,
relative log expression, normalized unscaled standard errors,
residual scale factors.


\section{Introduction}\label{I}

With the introduction of microarrays  biologist have been witnessing 
entire labs shrinking to matchbox size.  This paper invites quality 
researchers to join scientists on their {\it fantastic journey} into the 
world of microscopic high-throughput measurement technologies.  
Building a biological organism as laid out by the genetic code is a 
multi-step process with room for variation at each step. The first steps, 
as described by the {\it Dogma of molecular biology,} are genes (and 
DNA sequence in general), their transcripts and proteins.  Substantial factors 
contributing to their variation in both structure and abundance include 
cell type, developmental stage, genetic background and environmental conditions.  
Connecting molecular observations to the state of an organism is a 
central interest in molecular biology. This includes the study of the 
gene and protein functions and interactions, and their alteration in 
response to changes in environmental and developmental conditions.
Traditional methods in molecular biology generally work on a 
"one gene (or protein) in one experiment" basis.  With the invention 
of {\it microarrays} huge numbers of such macromolecules can 
now be monitored in one experiment. The most common kinds are 
{\it gene expression microarrays,} which measure the mRNA transcript
abundance for tens of thousands of genes simultaneously. 

For biologists, this high-throughput approach has opened up entirely
new avenues of research.  Rather than experimentally confirming the
hypothesized role of a certain candidate gene in a certain cellular 
process, they can use genome-wide comparisons to screen for all 
genes which might be involved in that process.  One of the first examples 
of such an exploratory approach is the expression profiling study of mitotic yeast cells
by \cite{CCWD98} which determined a set of a few 
hundred genes involved in the cell cycle and triggered a cascade of 
articles re-analyzing the data or replicating the experiment.  Microarrays 
have become a central tool in cancer research initiated by the discovery 
and re-definition of tumor subtypes based on molecular signatures 
(see e.g.~\cite{PSBB00}, \cite{AEBS00}, \cite{RG02}, \cite{yeoh02}).  
In Section \ref{B} we will explain different kinds of microarray technologies 
in more detail and describe their current applications in life sciences research.

A DNA microarray consists of a glass surface with a large number 
of distinct fragments of DNA called probes attached to it at fixed 
positions. A fluorescently labelled sample containing a mixture of 
unknown quantities of DNA molecules called the target is applied to the 
microarray. Under the right chemical conditions, single-stranded fragments of 
target DNA will base pair with the probes which are their complements,
with great specificity. This reaction is called hybridization, and is 
the reason DNA microarrays work. The fixed probes are either fragments 
of DNA called complementary DNA (cDNA) obtained from messenger RNA 
(mRNA), or short fragments known to be complementary to part of a gene, 
spotted onto the glass surface, or synthesized in situ. The point of 
the experiment is to quantify the abundance in the target of DNA 
complementary to each particular probe, and the hybridization reaction 
followed by scanning allows this to be done on a very large scale.  The 
raw data produced in a  microarray experiment consists of scanned images, 
where the image intensity in the region of a probe is proportional to the 
amount of labelled target DNA that base pairs with that probe. In this way 
we can measure the abundance of thousands of DNA fragments in a target 
sample. Microarrays based on cDNA or long oligonucleotide probes typically use 
just one or a few probes per gene.  The same probe sequence spotted in 
different locations, or probe sequences complementary to different parts 
of the same gene can be used to give within array replication. Short 
oligonucleotide microarrays typically use a larger number per gene, 
e.g.~11 for the HU133 Affymetrix array per gene. Such a set of 11 is called 
probeset for that gene, and the probes in a probe set are arranged 
randomly over the array. In the biological literature, microarrays are 
also referred to as (gene) chips or slides.

When the first microarray platforms were introduced in the early 90s, 
the most intriguing fact about them was the sheer number of genes
that could be assayed simultaneously.  Assays that used to be done 
one gene at a time, could suddenly be produced for thousands of genes 
at once.  A decade later, high-density microarrays would even fit entire genomes
of higher organisms.  
After the initial euphoria, the research community became 
aware that findings based solely on microarray measurements were 
not always as reproducible as they would have liked and that studies with inconclusive results 
were quite common. With this high-throughput measurement technology 
becoming established in many branches of life sciences research, 
scientists in both academic and corporate environments raised their 
expectations concerning the validity of the measurements.  Data quality
issues are now frequently addressed at meetings of the Microarray 
Gene Expression Database group (MGED).  The {\it Microarray Quality 
Control project,} a community-wide effort, under the auspices of the 
U.S. Food and Drug Administration (FDA), is aiming at establishing 
{\it operational metrics} to objectively assess the performance of 
seven microarray platform and develop minimal quality standards.
Their assessment is based on the performance of a set of standardized 
external RNA controls.  The first formal results of this project have been 
published in a series of articles in the September 2006 issue of {\it Nature 
Biotechnology.} 

Assessing the {\it quality of microarray data} has emerged as a new 
research topic for statisticians.  In this paper, we conceptualize
microarray data quality issues from a perspective which includes
the technology itself as well as their practical use by the research 
community.  We characterize the nature of microarray data from a 
quality assessment perspective, and we explain the different levels 
of microarray data quality assessment.  Then we focus on short 
oligonucleotide microarrays to develop a set of specific statistical 
data quality assessment methods including both numerical 
measures and spatial diagnostics.

Assumptions and hopes about the quality of the measurements have 
become a major issue in microarray purchasing.  Despite their 
substantially higher costs, Affymetrix short oligonucleotide microarrays
have become a widespread alternative to cDNA chips.  Informally, they are
considered the industrial standard among all microarray platforms. More recently,
Agilent's non-contact printed high-density cDNA microarrays and Illumina's
bead arrays have fueled the competition for high quality chips.  
Scientist feel the need for systematic quality assessment methods allowing 
them to compare different laboratories, different chip generation, or different platforms.  
They even lack good methods for selecting chips of good enough quality 
to be included in statistical data analysis beyond preprocessing.
We have observed several questionable practices in the recent past:
\begin{itemize}
\item Skipping hybridization QA/QC all together
\item Discarding entire batches of chips following the detection of
a few poor quality chips
\item Basing hybridization QA/QC on raw data rather than data that has
already been had large-scale technical biases removed
\item Delaying any QA/QC until all hybridizations are completed, thereby
losing the opportunity to remove specific causes of poor quality 
at an early stage
\item Focussing on validation by another measurement technolgy (e.g, quantitative PCR)
in Publication requirements rather than
addressing the quality of the microarray data in the first place
\item Merging of data of variable quality into one database with
the inherent risk of swamping it with poor quality data (as this 
produced  at a faster rate due to few replicates, less quality checks, 
less re-doing of failed hybridizations etc.)
\end{itemize} 

The community of microarray users has not yet agreed on a framework 
to measure accurary or precision in microarray experiments.  
Without universally accepted methods for quality assessment, 
and guidelines for acceptance, statisticians' judgements 
about data quality may be perceived as arbitrary by experimentalists.  
Users' expectations as to the level of gene expression data 
quality vary substantially.  They can depend on time frame and 
financial constraints, as well as on the purpose of their data collection. 
\cite{She39}, p.\,120/21, explained the standpoint of the applied scientist: \\
{\it ''He knows that if he were to
act upon the meagre evidence sometimes available to the pure scientist,
he would make the same mistakes as the pure scientist makes in estimates
of accuracy and precisions. He also knows that through his mistakes
someone may lose a lot of money or suffer physical injury or both.
[...] He does not consider his job simply that of doing the best he 
can with the available data; it is his job to get enough data before 
making this estimate.''}\\
Following this philosophy, microarray data used for medical diagnostics
should meet high quality standards. In contrast, microarray data collected 
for a study of the etiology of a complex genetic disease in a heterogeneous
population, one may decide to tolerate lower standards at the level of 
individual microarrays and invest the resources in a larger sample size.
Scientists need informative quality assessment tools to allow 
them to choose the most appropriate technology and optimal experimental
design for their precision needs, within their time and budget constraints.

The explicit goals of quality assessment for microarrays are manifold. 
Which goals can be envisioned depends on the resources and time
horizon and on the kind of user -- single small user, big user, 
core faculity, multi-center study, or "researcher into quality".  
The findings can be used to simply exclude chips from further study or 
recommend to have samples reprocessed.  They can be imbedded in a
larger data quality management and improvement plan.
Typical quality phenomena to look for include:
\begin{itemize}
\item Outlier chips 
\item Trends or patterns over time
\item Effects of particular hybridization conditions and sample characteristics
\item Changes in quality between batches of chips, 
cohorts of samples, lab sites etc.
\item Systematic quality differences between subgroups of a study
\end{itemize}

Some aspects of quality assessment and control for cDNA have been discussed in the literature. 
Among these, \cite{BFBV00} and \cite{FEGS02} emphasize the need for quality control 
and replication. \cite{WGG01} define a quality score for each spot based on intensity 
characteristics and spatial information, while \cite{hautaniemi03} approach
this with Baysian networks.  \cite{SYS03} and 
\cite{RDNS06} suggest explicit statistical quality measures based on individual 
spot observations using the image analysis software {\tt Spot} from \cite{YBS01}.
\cite{MKPA02} apply multivariate statistical process control to
detect single outlier chips.  The preprocessing and data management 
software package {\tt arrayMagic} of \cite{BHSP05} includes quality diagnostics.
The book by \cite{ZSA04} is a comprehensive collection 
of quality assessment and control issues concerning the various stages of 
cDNA microarray experiments including sample preparation, all from an experimentalist's 
perspective.  \cite{NoB05} suggest spot quality scores based on the variance of the
ratio estimates of replicates (on the same chip or on different chips).
Spatial biases have also been addressed.  
In examining the relationship between signal intensity and print-order, \cite{S02} 
reveals a plate-effect.  The normalization methodology by \cite{YDLS02} incorporates 
spatial information such as print-tip group or plate, to remove spatial biases created by
the technological processes. \cite{KQG03} and \cite{QKYG03} found pairwise 
correlations between genes due to their relative positioning of the spots on the slide 
and suggest a localized mean normalization method to adjust for this.  
\cite{tom05} proposed a method of identifying poor quality spots, and of addressing 
this by assigning quality weights.  \cite{RW05} developed an approach for the visualization 
and quantitation of regional bias applicable to both cDNA and Affymetrix microarrays.

For Affymetrix arrays, the commercial software \cite{AffyGCOS04} includes a 
{\it Quality report} with a dozen scores for each microarray (see  Subsection \ref{D_A}).
None of them makes use of the gene expression summaries directly, and   
there are no universally recognized guidelines as to which range should be 
considered good quality for each of the GCOS quality scores. 
Users of short oligonucleotide chips have found that the quality picture 
delivered by the GCOS quality report is incomplete or not sensitive enough, 
and that it is rarely helpful in assigning causes to poor quality.
The literature on quality assessment and control for short oligonucleotide arrays
is still sparse, though the importance of the topic has been stressed in numerous
places, and some authors have addressed have looked at specific issues.  
An algorithm for probeset quality assessment has been suggested by \cite{BolstadT}.
\cite{naef03} transfer the weight of a measurement to a subset of probes with
optimal linear response at a given concentration.  
\cite{gautier04} investigate the effect of updating the mapping of probes 
to genes on the estimated expression values.
\cite{SmH04} define four types of degenerate probe behaviour 
based on free energy computations and pattern recognition.  
\cite{finkelstein05} evaluated the Affymetrix quality reports
of over 5,000 chips collected by St.~Jude Children's Research Hospital over a 
period of three years, and linked some quality trends to experimental conditions.  
\cite{HGB05} extend traditional effect size models to combine data from different 
microarray experiments, incorporating a quality measure for each gene in each study.
The detection of specific quality issues such as the extraction, handling and amount 
of RNA, has been studied by several authors (e.g.\,\cite{archer05}, \cite{dumur04}, 
\cite{schoor03}, \cite{thach03}).

Before deriving new methods for assessing microarray data quality, we will relate 
the issue to established research into data quality from other academic disciplines,
emphasizing the particular characteristics of microarray data (Section \ref{D_C}). 
A conceptual approach to the statistical assessment of microarray data quality is 
suggested in Subsection \ref{D_A}, and is followed by a summary of the existing 
quality measures for Affymetrix chips.  

The theoretical basis of this paper is Section \ref{M}, where we 
introduce new numerical and spatial quality assessment methods for
short oligonucleotide arrays.  Two important aspects of our approach are:
\begin{itemize}
\item The quality measures are based on {\it all the data} from the array.
\item The quality measures are computed after hybridization and data preprocessing.
\end{itemize}
More specifically, we make use of probe level and probeset level quantities obtained  
as by-products of the Robust Multichip Analysis (RMA/fitPLM) preprocessing 
algorithm presented in \cite{IBCS03}, \cite{BCSS04} and \cite{GCBI04}.

Our {\it Quality Landscapes} serve as tools for visual quality inspection of the 
arrays after hybridization.
These are two dimensional pseudo-images of the chips based on probe level quantities,
namely the weights and residuals computed by RMA/fitPLM. These Quality 
Landscapes allow us to immediately relate quality to an actual location on the chip, 
a crucial step in detecting special causes for poor chip quality.  
Our numerical quality assessment is based on two distributions computed at
the probeset level, the {\it Normalized Unscaled Standard Error (NUSE)} 
and the {\it Relative Log Expression (RLE).}   Given a fairly general biological 
assumption is fulfilled, these distributions can be interpreted for chip 
quality assessment.  We further suggest ways of
conveniently visualizing and summarizing these distributions for larger chip sets
and of relating this quality assessment with other factors in the experiment, 
to permit the detection of special causes for poor quality to reveal biases.
Quality of gene expression data can be assessed on a number of levels, 
including that of probeset, chip and batch of chips.  
Another aspect of quality assessment concerns batches of chips. 
We introduce the {\it Residual Scale Factor (RSF),} a measure of chip batch quality. 
This allows us to compare quality across batches of chips within an experiment, 
or across experiments.
All our measures can be computed for all available types of short oligonuceotide 
chips given the raw data (CEL file) for each chip and the matching CDF file.
Software packages are described in \cite{BCBS05} and available at {\tt www.bioconductor.org}.

In Section \ref{R} we extensively illustrate and evaluate our quality 
assessment methods on the experimental microarray datasets described in Section \ref{S}.  
To reflect the fact that quality assessment is a necessary and fruitful 
step in studies of any kind, we use a variety of datasets, involving tissues 
ranging from fruit fly embryos to human brains, and from academic, 
clinical, and corporate labs.  We show how quality trends and 
patterns can be associated with sample characteristics and/or experimental 
conditions, and we compare our measures with the Affymetrix GCOS quality report.

\section{Background: Microarray technology and applications in 
biomedical research}\label{B}

After the hunt for new genes has dominated genetics in the 80s and 90s
of the last century, there has be a remarkable shift in molecular biology 
research goals towards a comprehensive understanding of the function 
of macromolecules on different levels in a biological organism. 
How and to what extend do genes control the construction and 
maintenance of the organism?  What is the role of intermediate gene 
products such as {\it RNA transcripts}? How do the macromolecules interact with others?
The latter may refer to {\it horizontal} interaction, such as genes with 
genes, or proteins with proteins.  It may also refer to {\it vertical} interaction, 
such as between genes and proteins.  {\it Genomics} and {\it proteomics} -- 
in professional slang summarized as {\it 'omics sciences} -- have started 
to put an emphasis on {\it functions.}  As the same time, these research 
areas have become more quantitative, and they have broadened the 
perspective in the sense of observing huge numbers of macromolecules 
simultaneously. These trends have been driven by recent biotechnological 
inventions, the most prominent ones being {\it microarrays.}  With these 
{\it high-throughput} molecular measurement instruments, the relative 
concentration of huge numbers of macromolecules can be obtained 
simultaneously in one experiment.  This section will give an overview 
of the biological background and the applications of microarrays in 
biomedical research.  For an extended introduction to {\it 'omics 
sciences} and to microarray-based research we refer to the excellent 
collections of articles  in the three Nature Genetics Supplements 
{\it The Chipping Forecast I, II, III} (1999, 2002, 2005)
and to the recent review paper by \cite{hoh06}.
\nocite{natureGeneticsSuppI}  \nocite{natureGeneticsSuppII} 
\nocite{natureGeneticsSuppIII} 

\subsection{Gene expression and construction of biological organisms}
\label{B-ge}

Though the popular belief about genes is still very deterministic --
once they are put into place, they function in a preprogrammed straight 
forward way -- for biologists the effect of a gene is variable.
Most cells in an organism contain essentially the same set of genes.
However, cells will look and act differently depending on which organ 
they belong to, the state of the organ (e.g.\,healthy vs.\,\,diseased), 
the developmental stage of the cell, or the phase of the cell cycle. 
This is predominantly the result of differences in the abundance, 
distribution, and state of the cells' proteins.  According to the 
{\it central dogma of molecular biology} the production of proteins is 
controlled by DNA (for simplicity, the exceptions to this rule are 
omitted here).  Proteins are polymers built up from 20 different kinds 
of amino acids.  Genes are {\it transcribed} into DNA-like macromolecules 
called {\it messenger RNA (mRNA)}, which goes from the chromosomes
to the {\it ribosomes.}  There,  {\it translation} takes place, converting
mRNA into the amino acid chains which fold into proteins.

The term {\it gene expression} is defined as the relative concentration 
of mRNA and protein produced by that gene.  Depending on the context, 
however, it is often used to refer to only one of the two.  The  {\it gene 
expression profile} of a type of cell usually refers to the relative abundance
of each of the mRNA species in the total cellular mRNA population.  
From a practical point of view, in particular by many areas of medical 
research, protein abundance is seen as generally more interesting than 
mRNA abundance. The measurement of protein abundances, however, 
is still much more difficult to measure on a large scale than mRNA abundance.  

\subsection{Microarray gene expression measurement
and applications}\label{B-ma}
There is one property which is peculiar to nucleic acids: their 
complementary structure.  DNA is reliably replicated by separating 
the two strands, and complementing each of the single strands 
to give a copy of the original DNA. The same mechanism can be 
used to detect a particular DNA or RNA sequence in a mixed sample.
The first tool to measure gene expression in a sample of cells of a 
was introduced in 1975.  A {\it Southern blot} -- named for its inventor -- 
is a multi-stage laboratory procedure which produces a pattern of 
bands representing the activity of a small set of pre-selected genes.  
During the 1980s spotted arrays on nylon holding bacterial colonies 
carrying different genomic inserts were introduced.  In the early 1990s, 
the latter would be exchanged for preidentified cDNAs.  The introduction 
of {\it gene expression microarrays} on glass slides in the mid 1990s 
brought a substantial increase in feature density.  With the new technology, 
gene expression measurements could be taken in parallel for thousands 
of genes.  Modern microarray platforms even assess the expression 
levels of tens of thousands of genes simultaneously. 

A gene expression microarray is a small piece of glass onto which 
{\it a priori} known DNA fragments called {\it probes} are attached at 
fixed positions.  In a chemical process called {\it hybridization,} the
microarray is brought into contact with material from a sample of cells.
Each probe binds to its complementary counterpart, an mRNA molecule 
(or a complementary DNA copy) from the sample, which we refer to as 
the {\it target}. The hybridization reaction product is made visible using 
fluorescent dyes or other (e.g.\,radioactive) markers, which are applied 
to the sample prior to hybridization.  The readout of the microarray 
experiment is a scanned image of the labelled DNA.  Microarrays are 
specially designed to interrogate the genomes of particular organisms, 
and so there are yeast, fruit fly, worm and human arrays, to name just a few. 

There are three major platforms for microarray-based
gene expression measurement: {\it spotted two-color cDNA arrays,}  
{\it long oligonucleotide arrays} and {\it short oligonucleotide arrays.}
In the platform specific parts of this paper we will focus on the latter.
On a short oligonucleotide microarray, each gene is represented on 
the array by a {\it probe set} that uniquely identifies the gene.
The individual probes in the set are chosen to have
relatively uniform hybridization characteristics. 
In the Affymetrix HU133 arrays, for example, 
each probe set  consists of 11 to 20 probe sequence pairs.  
Each pairs consists of a {\it perfect match (PM)} probe, a 25 bases long
oligonucleotide  that matches a part of the gene's sequence, 
and a corresponding {\it mismatch (MM)} probe, that has the same 
sequence as the PM except for the center base being flipped to its 
complementary letter.  The MM probes are intended to give an 
estimate of the random hybridization and cross hybridization signals, 
see \cite{LDBB96} and \cite{LFGL99} for more details.  
Other Affymetrix gene expression arrays may differ from the HU133 in
the number of probes per probe set.  Exon arrays do not have MM probes.  
Most of the arrays produced by Nimblegen are composed from 
60mer probes, but some are using 25mer probes. 
The number of probes per probeset is adapted to the
total number of probesets on the array to make optimal use of the space.
 
\subsection{Applications of gene expression microarrays in 
biomedical research} \label{B-ap}
Besides being more efficient than the classical gene-by-gene approach, 
microarrays open up entirely new avenues for research.  They offer
a comprehensive and cohesive approach to measuring the activity
of the genome.  In particular, this fosters the study of interactions.  
A typical goal of a microarray based research project is the search 
for genes that behave differently between different cell populations.
Some of the most common examples for comparisons are
diseased vs.~healthy cells, injured vs.~healthy tissue,  
young vs.~old organism, treated vs.~untreated cells.
More explicitly, life sciences researchers try to find answers 
to questions such as the following.  Which genes are affected by 
environmental changes or in response to a drug?
How do the gene expression levels differ across various mutants?
What is the gene expression signature of a particular disease?
Which genes are involved in each stage of a cellular process?
Which genes play a role in the development of an organism?
Or, more generally, which genes vary their activity with time?


\subsection{Other kinds of microarrays and their applications} \label{B-ot}
The principle of microarray measurement technology has been
used to assess molecules other than mRNA.  A number of platforms
are currently at various stages of development (see review by \cite{hoh06}).  
SNP chips detect single nucleotide polymorphisms.
They are an example for a well developed microarray-based 
genotyping platform.  CGH arrays are based on comparative 
genome hybridization. This method permits the analysis of changes in 
gene copy number for huge numbers of probes simultaneously.  
A recent modification, representational oligonucleotide microarray 
analysis (ROMA), offers substantially better resolution. Both SNP 
chips and CGH arrays are genome-based methods, which, in contrast 
to the gene expression-based methods, can exploit the stability of DNA. 
The most common application of these technologies is the localization 
of disease genes based on association with phenotypic traits.
Antibody protein chips are used to determine the level of proteins
in a sample by binding them to antibody probes immobilized on the 
microarray.  This technology is still considered semi-quantitative, 
as the different specificities and sensitivities of the antibodies can lead 
to an inhomogeneity between measurements that, so far, can not
be corrected for.  The applications of protein chips are similar to the 
ones of gene expression microarrays, except that the measurements
are taken one step further downstream.  More recent platforms 
address multiple levels at the same time.
ChIP-on-chip, also known as genome-wide location analysis, 
is a technique for isolation and identification of the DNA sequences 
occupied by specific DNA binding proteins in cells.

\subsection{Statistical challenges}\label{B-st}
The still growing list of statistical challenges stimulated by 
microarray data is a {\it tour d'horizon} in applied statistics;
see e.g.~\cite{Spe03}, \cite{mcl04} and \cite{wit04} for broad introductions.
From a statistical point of view a microarray experiment has three main challenges:
(i) measurement  process as multi-step biochemical and technological procedure
(array manufacturing, tissue acquisition, sample preparation, hybridization, scanning)
with each step contributing to the variation in the data; 
(ii) huge numbers measurements of different (correlated) molecular species
being take in parallel;
(iii) unavailability of 'gold-standards' covering a representative part of these species.
Statistical methodology has primarily been developed for gene expression 
microarrays, but most of the conceptual work applies directly to many kinds of
microarrays and many of the actual methods can be transferred to other 
microarray platforms fitting the characteristics listed above.  

The first steps of the data analysis, often referred to as {\it preprocessing}
or {\it low level analysis}, are the most platform-dependent tasks.  
For two-color cDNA arrays this includes image analysis (see e.g.~\cite{YBS01}) 
and normalization (see e.g.~\cite{YDLS02}). For short oligonucleotide chip
data this includes normalization (see e.g.~\cite{BIAS03}) and the estimation
of gene expression values (see e.g.~\cite{LiWo01} and \cite{IBCS03} as well as
subsequent papers by these groups). Questions around the design of micorarray 
experiments are mostly relevant for two-color platforms (see e.g.~Ch.~2 in 
\cite{Spe03}, \cite{Ker03} and further references there).

Analysis beyond the preprocessing steps is often referred to as {\it downstream 
analysis.}  The main goal is to identify genes which act differently in different 
types of samples.  Exploratory methods such as classification and cluster 
analysis have quickly gained popularity for microarray data analysis.  
For reviews on such methods from a statistical point of view see e.g.~Ch.~2 
and Ch.~3 in \cite{Spe03} and Ch.~3-7 in \cite{mcl04}.  On the other side of 
the spectrum, hypothesis-driven inferential statistical methods are now well
established and used. This approach typically takes a single-gene 
perspective in the sense that it searches for {\it individual} genes that are 
expressed differentially across changing conditions; see e.g.~\cite{DYSC02}.
The main challenge is the imprecision of the gene-specific variance estimate,
a problem that has been tackled by strategies incorporating a gene-unspecific
component into the estimate; see e.g.~\cite{ETST01}, \cite{LoS02}, \cite{CuiC05} 
and references therein, and \cite{taiS06} for the case of microarray time course data.  
Testing thousands of potentially highly correlated genes at 
the same time with only a few replicates raises a substantial multiple testing 
problem that has been systematically addressed by various authors incorporating 
Benyamini's and Hochberg's {\it false discovery rate (FDR)}; see e.g.~\cite{sto03} 
and the review \cite{DSB03}.
The joint analysis of pre-defined groups of genes based on {\it a priori} knowledge
has become an established alternative to the genome-wide exploratory approaches 
and the gene-by-gene analysis; see e.g.~\cite{gsea05} and \cite{bilN06}.

While methodology for microarray data analysis has become a fast growing
research area, the epistemological foundation of this research area shows gaps.
Among other issues, \cite{MTA04} addresses the problem of simultaneous validation 
of research results and research methods.  \cite{ACPS06} offer a review of the main 
approaches to microarray data analysis developed so far and attempt to unify them.
Many software packages for microarray data analysis have been made publicly 
available by academic researchers.  In particular, there is the BioConductor project, 
a community-wide effort to maintain a collection of R-packages for genomics
applications at {\tt www.bioconductor.org}.
Many of the main packages are described in \cite{BioCbook05}. 

\section{Microarrays and data quality}\label{D}

\subsection{Characteristics of high-throughput molecular data}\label{D_C}

Data quality is a well established aspect of many quantitative research fields.  
The most striking difference between assessing the quality 
of a measurement as opposed to assessing the quality of a manufactured 
item is the additional layer of uncertainty.  Concerns around the 
accuracy of measurements have a long tradition in physics and astronomy; 
the entire third chapter of the classical book \cite{She39}  
is devoted to this field.  Biometrics, psychometrics,
and econometrics developed around similar needs, and many academic
fields have grown a strong quantitative branch.  All of them facing
data quality questions.  Clinical trials is a field that is
increasingly aware of the quality of large data collections
(see \cite{GOKA95} and other papers in this special issue).  
With its recent massive move into the quantitative field, 
functional genomics gave birth to what some statisticians 
call {\it genometrics.}  We now touch on the major points that characterize 
gene expression microarray data from the point of view of QA/QC.
These points apply to other high-dimensional molecular measurements as well.

{\bf Unknown kind of data:} Being a new technology
in the still unknown terrain of functional genomics, microarrays
produce datasets with few known statistical properties, including shape of the 
distribution, magnitude and variance of the gene expression values, and the kind of
correlation between the expression levels of different genes.
This limits access to existing statistical methods.

{\bf Simultaneous measurements:}
Each microarray produces measurements for thousands of genes
simultaneously.  If we measured just one gene at a time, some version of
Shewhart control charts could no doubt monitor quality.  If we measured a small
number of genes, multivariate extensions of control charts might be adequate.  
In a way, the use of control genes is one attempt by biologists 
to scale down the task to a size that can be managed by these classical approaches.  
Control genes, however, cannot be regarded as typical representatives of 
the set of all the genes on the arrays. Gene expression measures are 
correlated because of both the biological interaction of genes, and 
dependencies caused by the common measurement process.  Biologically meaningful 
correlations between genes can potentially "contaminate" 
hybridization quality assessment.

{\bf Multidisciplinary teams:}
Microarray experiments are typically planned, conducted and evaluated by
a team which may include scientists, statisticians, technicians and physicians.
In the interdisciplinarity of the data production and handling,
they are similar to large datasets in other research areas.   
For survey data, \cite{G87} names the risk associated with 
such a {\it ``m{\'e}lange of workers''}.  Among other things,
he mentions: radically different purposes, lack of communication,
disagreements on the priorities among the components of quality and
concentration on the ``error of choice'' in their respective discipline.

The encouragement of a close cooperation between scientists and 
statisticians in the care for measurement quality
goes all the way back to \cite{She39}, p.70/71:
{\it ``Where does the statistician's work begin?
[...] before one turns over any sample of data to the statistician
for the purpose of setting tolerances he should first ask the
scientist (or engineer) to cooperate with the statistician in
examining the available evidence of statistical sontrol.  The
statistician's work solely as a statistician begins after the scientist
has satisfied himself through the application of control criteria that
the sample has arisen under statistically controlled conditions.''}

{\bf Systematic errors:}
As pointed out by \cite{Loe90}, and, in the context of
clinical trials, by \cite{MMBW84}, systematic errors 
in large datasets are much more relevant than random errors.
Microarrays are typically used in studies involving different 
experimental or observational groups. Quality differences between 
the groups are a potential source of confounding.  

{\bf Heterogenous quality in data collections:}
Often microarray data from different sources are merged into
one data collection.  This includes different batches of chips
within the same experiment, data from different laboratories
participating in a single collaborative study, or data from different
research teams sharing their measurements with the wider community.
Depending on the circumstances, the combination of data typically 
takes place on one of the following levels: raw data, preprocessed
data, gene expression summaries, lists of selected genes.
Typically, no quality measures are attached to the data.
Even if data are exchanged at the level of CEL files,
heterogeneity can cause problems.  Some laboratories filter out chips 
or reprocess the samples that were hybridized to chips that did not 
pass screening tests, others do not.  These are decision 
processes that ideally should take place according to the same criteria.
The nature of this problem is well known in data bank quality or 
{\it data warehousing} (see e.g.~\cite{WSF95}, \cite{WaR01}, \cite{ReT92}).   

{\bf Re-using of shared data:}
Gene expression data are usually generated and used to answer a particular
set of biological questions.    
Data are now often being placed on the web to enable the general community 
to verify the analysis and try alternative approaches to the original 
biological question.  Data may also find a secondary use in 
answering modified questions.  The shifted focus potentially requires
a new round of QA/QC, as precision needs might have changed and  
artifacts and biases that did not interfere with the original 
goals of the experiment may do so now.

{\bf Across-platform comparison:} 
\cite{She39}, p.~112, already values the consistency
between different measurement methods higher than consistency in repetition.  
For microarrays, consistency between the measurements of two or more platforms 
(two-color cDNA, long oligonucleotide, short oligonucleotide (Affymetrix), 
commercial cDNA (Agilent), and real-time PCR)
on RNA from the same sample has  been addressed in a number of publications.
Some of the earlier studies show little or no agreement  
(e.g.~\cite{KJB02}, \cite{ROSG03}, \cite{JHEM04}, \cite{ZPSB05}), while
others report mixed results (e.g.\,\cite{YWPS02}, \cite{BWH03}, \cite{WADC04}).
More recent studies improved the agreement between platforms by controlling for 
other factors. \cite{SSLA04} and \cite{YBWD04} restrict comparisons to subsets of 
genes above the noise level. \cite{mecham04} use sequence-based matching of 
probes instead of gene identifier-based matching. 
\cite{IWFY05}, \cite{WHBL05} and \cite{TRPS05} use superior preprocessing 
methods and systematically distinguish the lab effect from the platform effect;
see \cite{DKES06} and \cite{TRPS05} for detailed reviews and further references.
For Affymetrix arrays, \cite{WADC04}, \cite{DKBG05} and \cite{stevens05} 
found inter-laboratory differences to be managable.  However,
merging data from different generations of Affymetrix arrays is not as 
straightforward as one might expect (e.g.~\cite{nimgaonkar03}, \cite{MYBZ04}, 
\cite{mitchell04}, \cite{hwang04}, \cite{kong05}).
\subsection{Assessment of microarray data quality}\label{D_A}

Quality assessment for microarray data can be studied on at least seven levels:
\begin{itemize}
\item[(1)] the raw chip (pre-hybridization)
\item[(2)] the sample
\item[(3)] the experimental design
\item[(4)] the multi-step measurement process
\item[(5)] the raw data (post-hybridization)
\item[(6)] the statistically preprocessed microarray data 
\item[(7)] the microarray data as entries in a databank
\end{itemize}
The last two items are the main focus of this paper.  
The quality of the data after statistical processing 
(which includes background adjustment, normalization and 
probeset summarization) is greatly affected, but not entirely determined 
by the quality of the preceeding five aspects.

The raw microarray data (5) are the result of a multi-step procedure.
In the case of the expression microarrays this includes
converting mRNA in the sample to cDNA, labelling
the target mRNA via an in vitro transcription step, fragmenting and then
hybridizing the resulting cRNA to the chip, washing and staining, and
finally scanning the resulting array.
Temperature during storage and hybridization, the amount of sample and 
mixing during hubridization all have a substantial impact on the quality of 
the outcome.  Seen as a multi-step process (4) the quality management for
microarray experiments has a lot in common with chemical engineering, where
numerous interwoven quality indicators have to be integrated (see e.g.~\cite{MaY02}).
The designer of the experiment (3) aims to minimize 
the impact of additional experimental conditions (e.g.~hybridization date) 
and to maximize accuracy and precision for
the quantities having the hightest priority, given the primary objectives of the study.
Sample quality (2) is a topic in its own right, strongly tied to 
the organism and the institutional setting of the study.
The question how sample quality is related to the microarray data 
has been investigated in \cite{JGHH06} based on a variety of RNA quality 
measures and chip quality measures including both Affymetrix scores and 
and ours.  The chip before hybridization (1) is a manufactured item.
The classical theory of quality control for industrial mass production 
founded by \cite{She39} provides the appropriate framework for the 
assessment of the chip quality before hybridization.  

The Affymetrix software GCOS presents some chip-wide quality measures in the 
Expression Report (RTP file).  They can also be computed by the BioConductor R package
{\tt simpleaffy} described in \cite{wilson05}.  The document "QC and Affymetrix data" 
contained in this package discusses how these metrics can be applied.
The quantities listed below are the most commonly used ones from the Affymetrix report
(descriptions and guidelines from \cite{AffyGCOS04} and \cite{AffyQCGuidelines01}).
While some ranges for the values are suggested, 
the manuals mainly emphasize the importance of {\it consistency} of the measures within 
a set of jointly analyzed chips using similar samples and experimental conditions.  
The users are also encouraged to look at the scores in conjuction with others scores.
\begin{itemize}
\item {\bf Average Background:}  Average of the lowest 2$\%$ cell intensities on the chip.
Affymetrix does not issue official guidelines, but mentions that values typically range
from 20 to 100 for arrays scanned with the GeneChip Scanner 3000.
A high background indicates the presence of nonspecific binding of salts and cell debris 
to the array. 
\item {\bf Raw Q (Noise):}  Measure of the pixel-to-pixel variation of probe cells on
the chip.  The main factors contributing to Noise values are electrical noise of the scanner and
sample quality.  Older recommendations give a range of 1.5 to 3.  Newer sources, 
however, do not issue official guidelines because of the strong scanner
dependence.  They recommend that data acquired from the same scanner be checked for
comparability of Noise values.
\item {\bf Percent Present:}  
The percentage of probesets called {\it Present} by the Affymetrix detection algorithm.
This value depends on multiple factors including cell/tissue type, biological or environmental 
stimuli, probe array type, and overall quality of RNA. Replicate samples should have similar Percent Present values. Extremely low Percent Present values indicate poor sample quality. A general rule of thumb is human and mouse chips typically have 30-40 Percent Present, and yeast and E.~coli have 70-90 Percent Present. 
\item{\bf Scale Factor:}  Multiplicative factor applied to the signal values to make the
2$\%$ trimmed mean of signal values for selected probe sets equal to a constant.
For the HU133 chips, the default constant is 500.  No general recommendation
for an acceptable range is given, as the Scale Factors depend on the constant 
chosen for the scaling normalization (depending on user and chip type).
\item {\bf GAPDH 3' to 5' ratio (GAPDH 3'/5'):}  Ratio of the intensity of the 3' probe set to the 5' 
probe set for the gene GAPDH.  It is expected to be an indicator of RNA quality.
The value should not exceed 3 (for the 1-cycle assay).    
\end{itemize}

\section{Methods: A microarray quality assessment toolkit}
\label{M}
{\bf Perfect Match (PM):} 
The distribution of the (raw) PM values.  While we do not think 
of this as a full quality assessment measure, it can indicate
particular phenomena such as brightness or dimness of the image,
or saturation.  Using this tool in combination with other quality measures,
can help in detecting and excluding technological reasons for poor quality.
A convenient way to look at the PM distributions for a number of chips 
is to use boxplots.  Alternatively, the data can be summarized on the 
chip level by two single values:  the median of the PM of all probes on 
the chip, abbreviated {\it Med(PM),} and the interquartile range 
of the PM of all probes on the chip, denoted by {\it IQR(PM).} 

Our other assessment tools use probe level and probeset level quantities 
obtained as a by-product of the Robust Multichip Analysis (RMA) algorithm 
developed in \cite{IBCS03},  \cite{BCSS04} and \cite{GCBI04}.  
We now recall the basics about RMA and refer the 
reader to above papers for details.  Consider a fixed probeset.  
Let $y_{ij}$ denote the intensity of probe $j$ from this probeset on chip $i,$ 
usually already background corrected and normalized.  RMA is based on the model
\begin{equation}\label{RMAmodel}
\log_2 y_{ij} = \mu_i + \alpha_j + \varepsilon_{ij}, 
\end{equation}
with $\alpha_j$ a {\it probe affinity effect}, $\mu_i$ representing
the log scale expression level for chip $i,$ and $\varepsilon_{ij}$
an i.i.d.\ centered error with standard deviation $\sigma$.  For identifiability of the model,
we impose a zero-sum constraint on the $\alpha_j'$s. 
The number of probes in the probeset 
depends on the kind of chip (e.g.\,11 for the HU133 chip).  
For a fixed probeset, RMA robustly fits the model using iteratively weighted
least squares and delivers a probeset expression index $\hat{\mu}_i, $ for each chip. 

The analysis produces residuals $r_{ij}$ and weights $w_{ij}$ attached to probe $j$ 
on chip $i.$   The weights are used in the IRLS algorithm to achieve robustness.  
Probe intensities which are discordant with the rest of the probes in the set are 
deemed less reliable and downweighted.
The collective behaviour of all the weights (or all the residuals) on a chip
is our starting point in developing post-hybridization chip quality measures.
We begin with a "geographic" approach -- images of the chips that
highlight potential poorly performing probes -- and then continue with the
discussion of numerical quality assessment methods.

{\bf Quality landscapes:} 
An image of a hybridized chip can be constructed
by shading the positions in a rectangular grid according to
the magnitude of the perfect match in the corresponding position on the 
actual chip.  In the same way, the positions can be colored according to
probe-level quantities other than the simple intensities.  
A typical color code is to use shades of red for positive 
residuals and shades of blue for negative ones, with darker shades
corresponding to higher absolute values.  Shades of green are used 
for the weights, with darker shades indicating lower weights.  
As the weights are in a sense the reciprocals of the absolute residuals,
the overall information gained from these two types of quality landscapes is 
the same.  In some particular cases, the sign of the residuals 
can help to detect patterns that otherwise would have been overlooked
(see both fruit fly datasets in Sections \ref{R} for examples).

If no colors are available, gray level images are used.  
This has no further implications for the weight landscapes.  
For the residual landscapes, note that red and
blue shades are translated into similar gray levels, 
so the sign of the residuals is lost.  Positive and negative residuals 
can plotted on two separate images to avoid this problem.

{\bf Normalized Unscaled Standard Error (NUSE):} 
Fix a probeset.  Let $\hat{\sigma}$ be the estimated residual standard 
deviation in model (\ref{RMAmodel}) and 
$W_i=\sum_j w_{i j}$
the {\it total probe weight} (of the fixed probeset) in chip $i.$
The expression value estimate for the fixed probeset on chip $i,$
and its standard error are given by 
\begin{equation}\label{M:exprSum}
\hat{\mu}_i = \sum_j y_{ij}\,\cdot\,\frac{w_{ij}}{W_i} \quad\mbox{ and }\quad
SE(\hat\mu_i) = \frac{ \hat{\sigma} }{\sqrt{W_i}}.
\end{equation}

The residual standard deviations vary across the probesets within a chip.  
They provide an assessment of overall goodness of fit of the model to 
probeset data for all chips used to fit the model, but provide no 
information on the relative precision of estimated expressions across chips.  
The latter, however, is our main interest when we look 
into the quality of a chips compared to other chips in the same
experiment.  Replacing the ${\hat{\sigma}}$ by $1$ gives 
what we call the {\it Unscaled Standard Error (USE)} of the 
expression estimate.  Another source of heterogeneity is the number of 
``effective'' probes -- in the sense of being given substantial weight 
by the RMA fitting procedure.  That this number varies across probeset 
is obvious when different numbers of probes per probeset are used 
on the same chip.  Another reason is dysfunctional probes, 
that is, probes with high variabiliy, low affinity, or a tendency to crosshybridize.  
To compensate for this kind of heterogeneity, 
we divide the USE by its median over all chips and 
call this {\it Normalized Unscaled Standard Error (NUSE).}
\begin{equation}\label{M:NUSE}
NUSE(\hat\mu_i) = \frac{USE(\hat{\mu}_i)}{\mbox{Median}_\iota \{USE(\hat\mu_\iota)\}}=
\frac{1}{\sqrt{W_i}}\,\bigg/
\mbox{Median}_\iota\bigg\{\frac{1}{\sqrt{W_\iota}}\bigg\}.
\end{equation}
An alternative interpretation for the NUSE of a fixed probeset becomes 
apparent after some arithmetic manipulations. 
For any odd number of positive observations $a_\iota\ (\iota=1,...,I),$ 
we have $\mbox{Median}_\iota\{1/a_\iota\}=1/\mbox{Median}_\iota\, a_\iota,$
since the function $x\mapsto 1/x\ (x>0)$ is monotone.
For an even number $I,$ this identity is it still approximatively true.  
(The reason for the slight inaccuracy is that, for an even number, the
median is the average between the two data points in the center
positions.)  Now we can rewrite
\begin{equation}\label{M:NUSEaltern}
NUSE(\hat\mu_i)\approx
\frac{1}{\sqrt{W_i}}\,\bigg/
\frac{1}{\mbox{Median}_\iota\{\sqrt{W_\iota}\}}
=\frac{\mbox{Median}_\iota\{\sqrt{W_\iota}\}}{\sqrt{W_i}}
=\bigg(\frac{\sqrt{W_i}}{\mbox{Median}_\iota\{\sqrt{W_\iota}\}}\bigg)^{-1}.
\end{equation}
The total probe weight can also be thought of as an
{\it effective number of observations} contributing to the probeset
summary for this chip.  Its square root serves as the divisor in the 
standard error of the expression summaries \eqref{M:exprSum},
similarly to the role of $\sqrt{n}$ in the classical case
of the average of $n$ independent observations.
This analogy supposes, for heuristic purposes, that the probes are
independent;  in fact this is not true due to normalization,
probe overlap and other reasons.  The median of the total probe
weight over all chips serves as normalization constant.
In the form \eqref{M:NUSEaltern}, we can think of the NUSE as 
the reciprocal of the normalized square root of total probe weight.

The NUSE values fluctuate around 1.  Chip quality statements 
can be made based on the distribution of all the NUSE values of one chip.
As with the PM distributions, we can conveniently look at NUSE distributions 
as boxplots, or we can summarize the information on the chip level 
by two single values: 
The median of the NUSE over all probesets in a particular chip, 
{\it Med(NUSE),} and the interquartile range 
of the NUSE over all probesets in the chip, {\it IQR(NUSE).} 

{\bf Relative Log Expression (RLE):}
We first need a reference chip.  
This is typically the {\it median chip} which is constructed probeset by 
probeset as the median expression value over all chips in the experiment.  
(A computationally constructed reference chips such as this one is
sometimes called "virtual chip".)
To compute the RLE for a fixed probeset, take the 
difference of its log expression on the chip to its 
log expression on the reference chip.  Note that the RLE is not tied
to RMA, but can be computed from any expression value summary.
The RLE measures how much the measurement of the expression 
of a particular probeset in a chip deviates from measurements of the
same probeset in other chips of the experiment.   

Again, we can conveniently look at the distributions 
as boxplots, or we can summarize the information on the chip level 
by two single values: 
The median of the RLE over all probesets in a particular chip, 
{\it Med(RLE),} and the interquartile range 
of the RLE over all probesets in the chip, {\it IQR(RLE).} 
The latter is a measure of deviation of the chip from the median chip.
A priori this includes both biological and technical variability.  
In experiments where it can be assumed that 
\begin{equation}\label{M:BioAssumpA}
\text{\bf the majority of genes are not biologically effected,}
\end{equation}
IQR(RLE) is a measure of technical variability in that chip.  
Even if biological variability is present for most genes, 
IQR(RLE) is still a sensitive detector of sources of
technical variability that are larger than biological variability.
Med(RLE) is a measure of bias.
In many experiments there are reasons to believe that
\begin{equation}\label{M:BioAssumpB}
\text{\bf number of up regulated genes }\approx
\text{\bf number of down regulated genes. }
\end{equation}  
In that case, any deviation of Med(RLE) from
$0$ is an indicator of a bias caused by the technology.
The interpretation of the RLE depends on the assumptions
(\eqref{M:BioAssumpA} and \eqref{M:BioAssumpB})
on the biological variability in the dataset, but it provides
a measure that is constructed independently of the quality landscapes and the NUSE.

For quality assessment, we summarize and visualize
the NUSE, RLE, and PM distributions.  We found series of boxplots to 
be very a convenient way to glance over sets up to 100 chips.
Outlier chips as well as trends over time or pattern related to time 
can easily be spotted.  For the detection of systematic quality differences 
related to circumstances of the experiment, or to properties of the sample
it is helpful to color the boxes accordingly.  Typical coloring would be
according to groups of the experiment, sample cohort, lab site, hybridization 
date, time of the day, a property of the sample (e.g.\,time in freezer).  
To quickly review the quality of larger sets of chips, shorter summaries such 
as the above mentioned median or the interquartile range of PM, NUSE and 
RLE.  These single-value summaries at the chip level are also useful for 
comparing our quality measures to other chip quality scores in scatter plots, 
or for plotting our quality measures against continuous parameters related 
to the experiment or the sample.  Again, additional use of colors can draw 
attention to any systematic quality changes due to technical conditions. 

While the RLE is a form of absolute measure of quality, the NUSE is not.
The NUSE has no units.  It is designed to detect differences {\it between
chips within a batch.}  However, the magnitudes of these differences 
have no interpretation beyond the batch of chips analyzed together.
We now describe a way to attach a quality assessment to a set of chips as 
a whole.  It is based on a common residual factor for a batch of jointly 
analyzed chips, RMA estimates a common residual scale factor.  It enables
us to compare quality between different experiments, or between subgroups 
of chips in one experiment.  It has no meaning for {\it single} chips.

{\bf Residual scale factor (RSF):}  
This is a quality measure for batches of chips.  It does not apply to individual
chips, but assesses the quality of  batches of chips.  The batches can be 
a series of experiments or subgroups of one experiment (defined, e.g.\,by cohort, 
experimental conditions, sample properties, or diagnostic groups).  
To compute the RSF, assume the data are background corrected.  
As the background correction works on a chip by chip basis it does not 
matter if the computations were done simultaneously for all batches of chips 
or individually.  For the normalization, 
however, we need to find one target distribution to which we normalize 
all the chips in all the batches. This is important, since the target distribution
determines the scale of intensity measures being analyzed.  
We then fit the RMA model to each batch separately.  The algorithm delivers,
for each batch, a vector of the estimated {\it Residual Scales} for all the probesets.
We can now boxplot them to compare quality between batches of chips.
The median of each is called {\it Residual Scale Factor (RSF).}
A vector of residual scales is a heterogeneous set.  To remove the heterogeneity,
we can divide it, probeset by probeset, by the median over the estimated scales 
from all the batches.  This leads to alternative definitions of the quantities above,
which we call {\it Normalized Residual Scales} and 
{\it Normalized Residual Scale Factor (NRSF).}
The normalization leads to more discrimination between the batches, 
but has the drawback of having no units.  

Software for the computation and visualization of the quality measures
and the interpretation of the statistical plots is discussed in \cite{BCBS05}.
The code is publicly available from {\tt www.bioconductor.org}  
in the R package {\tt affyPLM}.
Note that the implementation of the NUSE in {\tt affyPLM}
differs slightly from the above formula.  It is based on the "true" standard error
as it is comes from M-estimation theory instead of the total weights expression 
in \ref{M:NUSE}. However, the difference is small enough not to matter for any 
of the applications the NUSE has in chip quality assessment.

\section{Datasets}\label{S}

{\bf Affymetrix HU95 spike-in experiments:}
Here 14 human cRNA fragments corresponding to transcripts known to be absent from RNA extracted from 
pancreas tissue were spiked into aliquots of the hybridization mix at different concentrations, which we call
chip-patterns.
The patterns of concentrations from the spike-in cRNA fragments across the chips form a Latin Square.
The chip-patterns are denoted by A, B,...,S and T, with A,...,L occurring just once, and M and Q being repeated 4 times each.
Chip patterns N, O and P are the same as that of M, while patterns R, S, and T are the same as Q.
Each chip-pattern was hybridized to 3 chips selected from 3 different lots referred to as the
L1521, the L1532, and the L2353 series.  See {\tt www.affymetrix.com/support/technical/sample.data/datasets.affx} for further details and data download.
For this paper, we are using the data from the 24 chips generated by chip patterns M, N, O, P, Q, R, S, T 
with 3 replicates each.

{\bf St.~Jude Children's Research Hospital leukemia data collection:} 
The study by \cite{yeoh02}  was conducted to determine whether gene expression 
profiling could enhance risk assignment for pediatric acute lymphoblastic leukemia (ALL).  
The risk of relapse plays a central role in tailoring therapy intensity.  A total of 389 samples were 
analyzed for the study, from which high quality gene expression data were obtained on 360 samples.
Distinct expression profiles identified each of the prognostically important leukemia subtypes, 
including T-ALL, E2A-PBX1, BCR-ABL, TEL-AML1, MLL rearrangement, and hyperdiploid$>$50 chromosomes.
In addition, another ALL subgroup was identified based on its unique expression profile.
\cite{ross03} re-analized 132 cases of pediatric ALL  from the original 327 
diagnostic bone marrow aspirates using the higher density U133A and B arrays. 
The selection of cases was based on having sufficient numbers of each subtype to build 
accurate class predictions, rather than reflecting the actual frequency of these groups in the 
pediatric population. The follow-up study identified additional marker genes for subtype discrimination,
and improved the diagnostic accuracy.
The data of these studies are publicly available as supplementary data.

{\bf Fruit fly mutant pilot study:}
Gene expression of nine fruit fly mutants were screened using
Affymetrix DrosGenome1 arrays.  The mutants are characterized by various forms of
dysfunctionality in their synapses. RNA was extracted from fly embryos,  pooled and labelled.
Three to four replicates per mutant were done.  Hybridization took place on six different days.
In most cases, technical replicates were hybridized on the same day.
The data were collected by Tiago Magalh\~aes in the Goodman Lab at the University of California, 
Berkeley, to gain experience with the new microarray technology.
 
{\bf Fruit fly time series:}
A large population of wild type (Canton-S) fruit flies was split into twelve cages
and allowed to lay eggs which were transferred into an incubator and aged
for 30 minutes.  From that time onwards, at the end of each hour for the next 12 hours, embryos 
from one plate were washed on the plate, dechorionated and frozen in liquid nitrogen.  
Three independent replicates were done for each time point.  
As each embryo sample contained a distribution of different ages, 
we examined the distribution of morphological stage-specific markers in each sample 
to correlate the time-course windows with the nonlinear scale of embryonic stages.
RNA was extracted, pooled, labeled and hybridized to
Affymetrix DrosGenome1 arrays.  Hybridization took place on two different days.
This dataset was collected by Pavel Toman{\u{c}}{\'{a}}k in the Rubin Lab at the University 
of California, Berkeley, as a part of their comprehensive study on spatial and temporal 
patterns of gene expression in fruit fly development \cite{tomancak02}.
The raw microarray data (.CEL files) are publically available at the project's website 
{\tt www.fruitfly.org/cgi-bin/ex/insitu.pl}.

{\bf  Pritzker data collection:}
The Pritzker neuropsychiatric research consortium uses
brains obtained at autopsy from the Orange Country Coroner's Office through the 
Brain Donor Program at the University of California, Irvine, Department of Psychiatry.
RNA samples are taken from the left sides of the brains.
Labeling of total RNA, chip hybridization, and scanning of oligonucleotide microarrays
are carried out at independent sites (University of California, Irvine;
University of California, Davis; University of Michigan, Ann Arbor). 
Hybridizations are done on HU95 and later generations of Affymetrix chips.
In this paper, we are looking at the quality of data used in two studies
by the Pritzker consortium.  The {\it Gender study} by \cite{vawter04} 
is motivated by gender difference in prevalence for some neuropsychiatric disorders.
The raw dataset has HU95 chip data on 13 subjects in three regions 
(anterior cingulate cortex, dorsolateral prefrontal cortex, and cortex of the cerebellar hemisphere).
The {\it Mood disorder study} described in \cite{bunney03} is based on a 
growing collection of gene expression measurements in, ultimately, 25 regions. 
Each sample was prepared and then split so that it could be hybridized to the chips in both 
Michigan and either Irvine or Davis.
 
\section{Results}\label{R}

We start by illustrating our quality assessment methods on the
well known Affymetrix spike-in experiments.  The quality of these chips
is well above what can be expected from an average lab experiment.
We then proceed with data collected in scientific studies from variety of 
tissue types and experimental designs.  Different aspects of quality analysis 
methods will be highlighted throughout this section.  Our quality analysis results 
will be compared with the Affymetrix quality report for several sections of the 
large publicly available St.~Jude Children's Research Hospital 
gene expression data collection. 

(A) {\bf Outlier in the Affymetrix spike-in experiments:}
24 HU 95A chips from the Affymetrix spike-in dataset.
All but the spike-in probesets are expected to be non-differentially
expressed across the arrays.  As there are only 14 spike-ins out
of about twenty thousand probesets, they are, from the quality 
assessment point of view, essentially 24 identical hybridizations.  
A glance at the weight (or residual) landscapes 
gives a picture of homogenous hybridizations
with almost no local defects on any chip but \#20 (Fig.~A1).  
The NUSE indicates that chip \#20 is an outlier.  Its median is
well above 1.10, while all others are smaller than 1.05,
and its IQR is three and more times bigger than it is for any other chip (Fig.~A2).  
The series of boxplots of the RLE distributions confirms these findings.  
The median is well below $0,$ and the IQR is two and more times bigger than 
it is for any other chip. Chip \#20 has both a technologically caused bias and 
a higher noise level.  The Affymetrix quality report (Fig.~A3), however, does 
not clearly classify \#20 as an outlier.  Its GAPDH 3'/5'
of about 2.8 is the largest within this chip set, but the value 2.8 is considered
to be acceptable.  According to all other Affymetrix quality measures 
-- Percent Present, Noise, Background Average, Scale Factor -- 
chip \#20 is within a group of lower quality chips, but does not stand out.

(B) {\bf Outlier in St.~Jude's data not detected by the 
Affymetrix quality report:}  The collection of MLL HU133B chips consists of 
20 chips one of which turns out to be an outlier.  The NUSE boxplots 
(Fig.~B1, bottom line) show a median over 1.2 for chip \#15 while all others 
are below 1.025.  The IQR is much larger for chip \#15 than it is for any other chip.
The RLE boxplots (Fig.~B1, top line) as well distinguish chip \#15 as an obvious outlier.
The median is about $-0.2$ for the outlier chip, while it is very close to $0$ for 
all other chips.  The IQR is about twice as big as the largest of the IQR of the other 
chips.  Fig.~B1 displays the weight landscapes of chip\#15 along with those of
two of the typical chips.  A region on the left side of chip \#15,
covering almost a third of the total area, is strongly down weighted,
and the chip has elevated weights overall.  
Affymetrix quality report (Fig. B3) paints a very different picture --  Chip $\#$15 is an
outlier on the Med(NUSE) scale, but does not stand out on any of common Affymetrix
quality assessment measures: Percent Present, Noise, Scale Factor, and GAPDH 3'/5'.

(C) {\bf Overall comparison of our measures and the Affymetrix  
quality report for a large number of St.~Jude's chips:}
Fig.~D1 pairs the Med(NUSE) with the four most common GCOS scores
on a set of 129 HU133A chips from the St.~Jude dataset.
There is noticable linear association between Med(NUSE) and 
Percent Present, as well as between Med(NUSE) and Scale Factor.
GAPDH 3'/5' does not show a linear association with any the other scores.

(D) {\bf Disagreement between our quality measures and the Affymetrix 
quality report for Hyperdip$>$50 subgroup in St.~Jude's data:}  
The Affymetrix quality report detects problems with many 
chips in this dataset.  For chip A,
Raw Q (Noise) is out of the recommended range for 
the majority of the chips: $\#$12, $\#$14, C1, C13, C15, C16, C18, 
C21, C22, C23, C8 and R4. 
Background detects chip $\#$12 as an outlier. 
Scale Factor does not show any clear outliers.  
Percent Present is within the typical range for all chips.  
GAPDH 3'/5' is below 3 for all chips.
For chip B, Raw Q (Noise) is out of the recommended range for 
the $\#$12, $\#$8, $\#$18 and R4.
Background detects chip $\#$12 and $\#$8 as outliers. 
Scale Factor does not show any clear outliers.  
Percent Present never exceeds 23$\%$ in this chip set,
and it is below the typical minimum of 20$\%$ for chips
$\#$8, C15, C16, C18, C21 and C4.  GAPDH  3'/5' is 
satisfactory for all chips.

Our measures suggest that, with one exception, the chips are 
of good quality (Fig.~D1).  The heterogeneity of the perfect match
distributions does not persist after the preprocessing.
For chip A, $\#$12 has the largest IQR(RLE) and is a clear
outlier among the NUSE distributions.  Two other chips have
elevated IQR(RLE), but do not stand out according to NUSE.
For chip B, the RLE distributions are very similar with
$\#$12 again having the largest IQR(RLE).  The NUSE 
distributions are consistently showing good quality with 
the exception of chip $\#$12.

(E) {\bf Varying quality between diagnostic subgroups in the St.~Jude's data:}
Each boxplot in Fig.~E1 sketches the Residual Scale Factors (RSF) of the chips
of all diagnostic subgroups.  They show substantial quality differences.  
The E2A$\_$PBX1 subgroup has a much higher Med(RSF) than the other subgroups. 
The T$\_$ALL subgroup has a slightly elevated Med(RSF) and a higher IQR(RSF) than
the other subgroups.

(F) {\bf Hybridization date effects on quality of fruit fly chips:}
The fruit fly mutant with dysfunctional synapses is an experiment of the
earlier stages of working with Affymetrix chips in this lab.  It shows a wide
range of quality.  In the boxplot series of RLE and NUSE (Fig.~F1) a dependency of
the hybridization date is striking.  The chips of the two mutants hybridized on the day 
colored yellow show substantially lower quality than any of the other chips.
Fig.~F2 shows a weight landscape revealing smooth mountains and valleys.
While the pattern is particularly strong in the chip chosen for this picture, 
it is quite typical for the chips in this dataset.  We are not sure about the 
specific technical reason for this, but assume it is related to insufficient mixing 
during the hybridization.

(G) {\bf Temporal trends or biological variation in fruit fly time series:}  
The series consists of 12 developmental stages of fruit fly embryos
hybridized in 3 technical replicates each.  While the $log_2$(PM) distributions 
are very similar in all chips, we can spot two kinds of systematic patterns in the RLE 
and NUSE boxplots (Fig.~G1).  One pattern is connected to the developmental stage.
Within each single one of the three repeat time series, the hybridizations
in the middle stages look "better" than the ones in the early stages and 
the chips in the late stages.  This may, at least to some extent, be due to 
biological rather than technological variation.  In embryo development, 
especially in the beginning and at the end, huge numbers of genes are 
expected to be affected, which is a potential violation of 
assumption \eqref{M:BioAssumpA}.  Insufficient staging in the first
very short developmental stages may further increase the variability.  
Also, in the early and late stages of development, there is substantial doubt 
about the symmetry assumption \eqref{M:BioAssumpB}. 
Another systematic trend in this dataset is connected to the repeat series.  
The second dozen chips are of poorer quality than the others.
In fact, we learned that they were hybridized on a different day from the rest.

The pairplot in Fig.~G2 looks at the relationship between our chip quality measures.
There is no linear association between the raw intensities -- summarized as
Med(PM) -- and any of the quality measures. A weak linear association can
be noted between Med(RLE) and IQR(RLE).  It is worth to note that is 
becomes much stronger when focusing on just the chips hybridized on
the day colored in black.  IQR(RLE) and Med(NUSE) again have a weak 
linear association which becomes stronger when looking only at one of the 
subgroups, except this time it is the chips colored in gray.
For the pairing Med(RLE) and Med(NUSE), however, there is no 
linear relationship.  Finally (not shown), as in the dysfunctional 
synapses mutant fruit fly dataset, a double-wave gradient, as seen in Fig.~F2 
for the other fruit fly dataset, can be observed in the quality landscapes of 
many of the chips.  Although these experiments were conducted by a 
different team of researchers, they used the same equipment as 
that used in generating the other fruit fly dataset.

(H) {\bf Lab differences in Pritzker's gender study:}
We looked at HU95 chip data from 13 individuals in two brain regions,
the cortex of the cerebellar hemisphere (short: cerebellum) 
and the dorsolateral prefrontal cortex.
With some exceptions, each sample is hybridized in both lab M and lab I.  
The NUSE and RLE boxplots (Fig.~H1) for the cerebellum dataset display 
an eye-catching pattern:
They show systematically much better quality in Lab M then in Lab I.  
This might be caused by overexposure or saturation effects in Lab I.  
The medians of the raw intensities (PM) values in Lab I are, on a 
$log_2$-scale between about 9 and 10.5, while they are 
very consistently about 2 two 3 points lower in Lab M.  
The dorsolateral prefrontal cortex hybridizations show, for the most part, a lab effect similar to the 
one we saw in the cerebellum chips (plots not shown here).

(I) {\bf Lab differences in Pritzker's mood disorder study:} 
After the experiences with lab differences in the gender study, 
the consortium went through extended efforts to minimize these problems.  
In particular, the machines were calibrated by Affymetrix specialists.
Fig.~I1 summarizes the quality assessments of three of the Pritzker
mood disorder datasets.  We are looking at HU95 chips from 
two sample cohorts (a total of about 40 subjects) in each of the brain regions
anterior cingulate cortex, cerebellum, and dorsolateral
prefrontal cortex.  In terms of Med(PM), for each of the three 
brain regions, the two replicates came closer to each other:
the difference between the two labs in the mood disorder study is a 
third or less of the difference between the two labs in the gender study
(see first two boxes in each of the three parts of Fig.~I1, and compare with Fig.~H1). 
This is due to lab I dropping in intensity (toward lab M) and the new lab D 
also operating at that level.  The consequence of the intensity adjustments
for chip quality do not form a coherent story.  While for cerebellum the quality in   
lab M is still better than in the replicate in one of the other labs, for the other two
brain regions the ranking is reversed. Effects of a slight underexposure in lab M
may now have become more visible.  Generally, in all brain regions, 
the quality differences between the two labs are still there, but they
are much smaller than they in the gender study data.

(J) {\bf Assigning special causes of poor quality for St.~Jude's data:}
Eight quality landscapes from the early St.~Jude's data, a collection of 335 
HU133Av2 chips.  The examples were picked for being particularly strong
cases of certain kinds of shortcomings that repeatedly occur in this chip collection.  
They do not represent the general quality level in the early St.~Jude's chips,
and even less so the quality of later St.~Jude's chips.  
The figures in this paper are in gray levels.
If the positive residual landscape is shown, the negative residual landscape
is typically some sort of complementary image, and vice versa.  
Colored quality landscapes for all St.~Jude's chips can be downloaded from 
Bolstad's {\it Chip Gallery} at {\tt www.plmimagegallery.bmbolstad.com}. 

Fig.~J1 "Bubbles" is the positive residual landscape of chip Hyperdip-50-02.  
There are small dots in the left upper part of the slide,  and two bigger 
ones in the middle of the slide.  We attribute the dots to dust attached 
to the slide or air bubbles stuck in this place during the hybridization.
Further, there is an accumulation of positive residuals in the bottom
right corner.  Areas of elevated residuals near the corners and edges 
of the slide are very common, often much larger than in this chip.  
Mostly they are positive. The most likely explanation are air bubbles that, 
due to insufficient mixing during the hybridization, got stuck close to the 
edges where they had gotten when this edges was in a higher position
to start with or brought up there by the rotation.  Note that there 
typically is some air in the solution injected into the chip (through a
little hole near one of the edges), but that the air is moved around
by the rotation during the hybridization to minimize the effects on
the probe measurements.

Fig.~J2 "Circle and stick" is the residual landscape of Hyperdip47-50-C17.  
This demonstrates two kinds of spatial patterns that are probably caused by
independent technical shortcomings.  First, a circle with equally spaced darker 
spots (approximately).  The symmetry of the shape suggests it was caused 
by a foreign object scratching trajectories of the rotation during the hybridization 
into the slide.  Second, there are little dots that almost seem to be aligned along 
a straight line connecting the circle to the upper right corner.  The dots might be 
air bubbles stuck to some invisible thin straight object or scratch.

Fig.~J3 "Sunset" is the negative residual landscape of Hyperdip-50-C6.
This chip illustrates two independent technical deficiencies.  First, there is
a dark disk in the center of the slide.  It might be caused by insufficient mixing, 
but the sharpness with which the disk is separated from the rest of the image
asks for additional explanations.  Second, the image obviously splits
into an upper and a lower rectangular part with different residual, 
separated by a straight border.  As a most likely explanation, we attribute
this to scanner problems.

Fig.~J4 "Pond" is the negative residual landscape of TEL-AML1-2M03.
The nearly centered disc covers almost the entire slide.  It might be caused
by the same mechanisms that were responsible for the smaller disc in the 
previous figure.  However, in this data collection, we have only seen 
two sizes of discs -- the small disk as in the previous figure and the large
one as in this figure.  This raises the question why the mechanism that causes
them does not produce medium size discs. 

Fig.~J5 "Letter S" is the positive residual landscape of Hypodip-2M03. 
The striking pattern -- the letter 'S' with the "cloud" on top --
is a particularly curious example of a common technical shortcoming.
We attribute the spatially heterogeneous  distribution of the residuals to
insufficient mixing of the solution during the hybridization.

Fig.~J6 "Compartments" is the positive residual landscape of Hyperdip-50-2M02.
This is a unique chip.  One explanation would be that the vertical
curves separating the three compartments of this image are long thin
foreign objects (e.g.~hair) that got onto the chip and blocked
or inhabited the liquid from being spread equally over the entire chip.

Fig.~J7 "Triangle" is the positive residual landscape of TEL-AML1-06.
The triangle might be caused by a long foreign object stuck to
the center of the slide on one end and free, and hence manipulated
by the rotation, on the other end.

Fig.~J8 "Fingerprint" is the positive residual landscape of Hyperdip-50-C10.
What looks like a fingerprint on the picture might actually be one.
With the slide measuring 1cm by 1cm, the pattern has about the size of
a human fingerprint or the middle part of it.

\section{Discussion}

{\bf Quality landscapes:} 
The pair of the positive and negative residual landscapes contains the 
maximum information.  Often, one of the two residual landscapes can 
already characterize most of the spatial quality issues.
In the weight pictures, the magnitude of the derivation is preserved,
but the sign is lost.  Therefore, unrelated local defects can appear
indistinguishable in weight landscapes.
The landscapes allow a first glance at the overall quality of the array:
A square filled with low-level noise typically comes from a good quality 
chip, one filled with high-level noise comes from a chip with uniformly 
bad probes.  If the landscape is reveals any spatial patterns, the quality may 
or may not be compromised depending on the size of the problematic area.
Even a couple of strong local defects may not lower the chip quality,
as indicated by our measures.  
The reason lies in both the chip design and the RMA model.  The probes 
belonging to one probeset are scattered around the chip assuring that a 
bubble or little scratch would only affect a small number of the probes in 
a probeset; even a larger under- or overexposed area of the chip may 
affect only a minority of probes in each probeset.  
As the RMA model is fitted robustly, its expression summaries are
shielded against this kind of disturbance.

We found the quality landscape most useful in assigning special causes
of poor chip quality.  A quality landscape composed of smooth mountains 
and valleys is most likely caused by insufficient mixing during the hybridization.  
Smaller and sharper cut-out areas of elevated residuals are typically related to
foreign objects (dust, hair, etc.) or air bubbles.  Symmetries can indicate
that scratching was caused by particles being rotated during hybridization. 
Patterns involving horizontal lines may be caused by scanner miscalibration.  
It has to be noted, that the above assignment of causes are educated guesses
rather than facts.  They are the result of extensive discussions with experimentalist, 
but there remains a speculative component to them.  Even more hypothetical
are some ideas we have regarding how the sign of the residual could
reveal more about the special cause.  All we can say at this point is that 
the background corrected and normalized probe intensity deviate from what
the fitted RMA model would expect them to be.
The focus in this paper is a global one: chip quality.
Several authors have worked on spatial chip images from a different perspective,
that of automatically detecting and describing local defects 
(see \cite{RW05}, or the R-package {\tt Harshlight} by \cite{suarez05}).
It remains an open question how to use this kind of assessment
beyond the detection and classification of quality problems. In our approach, if we do not 
see any indication of quality landscpape features in another quality
indicator such as NUSE or RLE, we suppose that it has been rendered harmless
by our robust analysis. This may not be true.

{\bf RLE:}  Despite its simplicity the RLE distribution 
turns out to be a powerful quality tool.  For a small number of chips,
boxplots of the RLE distributions of each chip allow the detection of
outliers and temporal trends.  The use of colors or gray levels for 
different experimental conditions or sample properties facilitates the 
detection of more complex patterns.  For a large number of chips, 
the IQR(RLE) is a convenient and informative univariate summary.  
Med(RLE) should be monitored as well to detect bias.
As seen in the Drosophila embryo data, these assumptions are crucial
to ensuring that what the RLE suggests really are technical artifacts 
rather than biological differences.  Note that the RLE is not tied 
to the RMA model, but could as well be computed based on expression 
values derived from other algorithms.  The results may differ, but our 
experience is that the quality message turns out to be similar. 

{\bf NUSE:}  As in the case of the RLE, boxplots of NUSE distributions
can be used for small chip sets, and plots of their median and interquartile 
ranges serve as a less space consuming alternative for larger chip sets.
For the NUSE, however, we often observe a very high correlation
between median and interquartile range, so keeping track of just
the median will typically suffice.  Again, colors or gray levels can 
be used to indicate experimental conditions facilitating the detection their 
potential input on quality.  The NUSE is the most sensitive of our quality 
tools, and it does not have a scale.  Observed quality differences,
even systematic ones, have therefore to carefully assessed.  
Even large relative differences do not necessarily compromise an experiment, 
or render useless batches of chips within an experiment.  They should always alert the 
user to substantial heterogeneity, whose cause needs to be investigated. 
On this matter we repeat an obvious but important principle we apply. When there
is uncertainty about whether or not to include a chip or set of chips in an
analysis, we can do both analyses and compare the results. If no great differences
result, then sticking with the larger set seems justifiable.

{\bf Raw intensities and quality measures:}  Raw intensities are not 
useful for quality prediction by itself, but they can provide some
explanation for the poor performance according to the other quality measures.
All the Pritzker datasets, for example, suffer from systematic differences
in the location of the PM intensity distribution (indicated by Med(PM)).
Sometimes the lower location was worse -- too close to underexposure --
and sometimes the higher was worse -- too close to overexposure or saturation.
We have seen examples of chips for which the raw data give misleading 
quality assessment.  Some kinds of technological shortcomings can be 
removed without trace by the statististical processing, while others remain.

{\bf Comparison with Affymetrix quality scores:}
We found good overall agreement between our quality assessment and two of the
Affymetrix scores:  Percent Present and Scale Factor.  Provided the quality in a chips set 
covers a wide enough range, we typically see at least a weak linear association
between our quality measures and these two, and sometimes other Affymetrix quality scores.
However, our quality assessment does not always agree with the Affymetrix quality report.
In the St.~Jude data collection we saw that the sensitivity of the Affymetrix quality report 
could be insufficient.  While our quality assessment based on RLE and NUSE clearly 
detected the outlier chip in the MLL chip B dataset, none of the measures in the Affymetrix 
quality did.  The reverse situation occured in the Hyperdip chip A dataset.
While most of the chips passed according to our quality measures, 
most of the chips got a poor Affymetrix quality scores.  

{\bf RSF:}  The Residual Scale Factor can detect quality differences between
parts of a data collection, such as the diagnostic subgroups in the St.~Jude's
data.  In the same way, it can be employed to investigate quality differences
between other dataset divisions defined by sample properties, lab site, scanner, 
hybridization day, or any other experimental condition.  More experience as
to what magnitudes of differences are acceptable is still needed.

\section{Conclusions and Future Work}

In this paper, we have laid out a conceptual framework for a statistical approach 
for the assessment and control of microarray data quality. In particular,
we have introduced a quality assessment toolkit for short oligonucleotide arrays.  
The tools highlight different aspects in the wide spectrum of potential quality 
problems.  Our numerical quality measures, the NUSE and the RLE,  are an 
efficient way to detect chips of unusually poor quality.  
Furthermore, they permit the detection of temporal trends and patterns, batch effects, 
and quality biases related to sample properties or to experimental conditions.  
Our spatial quality methods, the weight and residual landscapes, add to the 
understanding of specific causes of poor quality by marking those regions 
on the chip where defects occur.  Furthermore, they illustrate the robustness
of the RMA algorithm to small local defects.  The RSF quantifies quality differences
between batches of chips.  It provides a broader framework for the quality scores
of the individual chips in an experiment.  
All the quality measures proposed in this paper can be computed based on the 
raw data using publicly available software. 

Deriving the quality assessment directly from the statistical model used to 
compute the expression values is more powerful than basing it on the 
performance of a particular set of of control probes, because the control probes 
may not behave in a way that is representative for the whole set of probes 
on the array.  The model-based approach is also preferable to 
metrics less directly related to the bulk of the expression values.
Some of the Affymetrix metrics, for example, are derived from the raw probe
values and interpret any artifacts as quality problems, even if they are
removed by routine preprocessing steps. A lesson from the practical examples 
in this paper is the importance of a well designed experiment.  
One of the most typical sources for bias, for example, is an unfortunate
systematic connection between hybridization date and groups of the study --
a link that could have been avoided by better planning.

More research needs to be done on the attribution of specific causes to poor quality
measurements.
While our quality measures, and, most of all, our quality landscapes, are a rich 
source for finding specific causes of poor quality, a speculative component remains.
To increase the credibility of the diagnoses, 
systematic quality experiments need to be conducted.  
A big step forward is Bolstad's {\it Chip Gallery} at {\tt www.plmimagegallery.bmbolstad.com,} 
which collects quality landscapes from Affymetrix chip collections, provides details 
about the experiment and sometimes offers explanations for the technical 
causes of poor quality.  Started as a collection of chip curiosities this website
is now growing into a visual encyclopedia for quality assessment.
Contributions to the collection, in particular those with known causes
of defects, are invited (anonymous if preferred). 
Further methodological research is needed to explore the use of the spatial 
quality for statistically "repairing" local defects, or making partial use
of locally damaged chips.

We are well aware that the range of acceptable values for each quality 
scores is the burning question for experimentalists.  
Our quality analysis results with microarray datasets from a variety of scientific 
studies in Section \ref{R} show that the question about the right threshold for 
good chip quality does not have a simple answer yet, at least not as the present 
level of generality.  
Thresholds computed for gene expression measurements in fruit fly mutant 
screenings can not necessarily be transferred to brain disease research or to 
leukemia diagnosis.  Thresholds need to be calibrated to the tissue type, 
the design, and the precision needs of the field of application.  
We offer two strategies to deal with this on different levels: 
\begin{enumerate}
\item {\bf Individual researchers:} We encouraged researchers to look for 
outliers and artificial patterns in the series of quality measures of the 
batch of jointly analyzed chips.  Furthermore, any other form of unusual 
observations -- e.g.\,a systematic disagreement between NUSE and RLE, 
or inconsistencies in the association between raw intensities and quality 
measures -- potentially hints at a quality problem in the experiment.
\item {\bf Community of microarray users:}  We recommend the development
of quality guidelines.  They should be rooted in extended collections of
datasets from scientific experiments.  Complete raw datasets are ideal, 
where no prior quality screening has been employed.  Careful 
documentation of the experimental conditions and properties help to
link unusual patterns in the quality measures to specific causes.  
The sharing of unfiltered raw chip data from scientific experiments on the 
web and the inclusion of chip quality scores in gene expression databank 
entries can lead the way towards community-wide quality standards.
Besides, it contributes to a better understanding how quality measures 
relate to special causes of poor quality.  
In addition, we encourage the conduction of {\it designed microarray quality 
experiments.}  Such experiments aim at an understanding of the effects of
RNA amount, experimental conditions, sample properties and sample
handling on the quality measures as well as on the downstream analysis.
They give an idea of the range of chip quality to be expected under 
given certain experimental, and, again, they help to characterize specific 
causes of poor quality.
\end{enumerate}

Benchmarking of microarray data quality will happen one way or another;
if it is not established by community-wide agreements the individual 
experimentalist will resort to judging on the basis of anecdotal evidence.  
We recommend that benchmarks be actively developed by the community of 
microarray researchers, experimentalists and data analysts. 
The statistical concepts and methods proposed here may serve as a
foundation for the quality benchmarking process.

\section{Acknowledgement}
We thank St.~Jude Children's Research Hospital, the Pritzker Consortium, 
Tiago Magalh\~aes, Pavel Toman{\u{c}}{\'{a}}k and Affymetrix for sharing their data 
for quality assessment purposes.

\addcontentsline{toc}{chapter}{\numberline{}Bibliography}
\bibliographystyle{asa}




\vfill\newpage

\section*{Figures}


\renewcommand{\thefigure}{A1}
\begin{figure*}[htbp]
\begin{center}
\includegraphics[width=6cm]{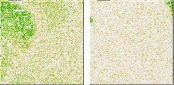}
\caption
{{\em
Weight landscapes of two chips from the Affymetrix HU95 spike-in experiment.   The chip on the right-hand side is typical for this data set, only tiny local defects and an overall good probe performance.  The chip of the left-hand side (chip $\#$20) has more down-weighted probes all over the slide, and it has a large down-weighted area in the upper left part of the slide\hspace{-1.3mm}
}}
\end{center}
\end{figure*}

\renewcommand{\thefigure}{A2}
\begin{figure*}[htbp]
\begin{center}
   \includegraphics[width=14cm]{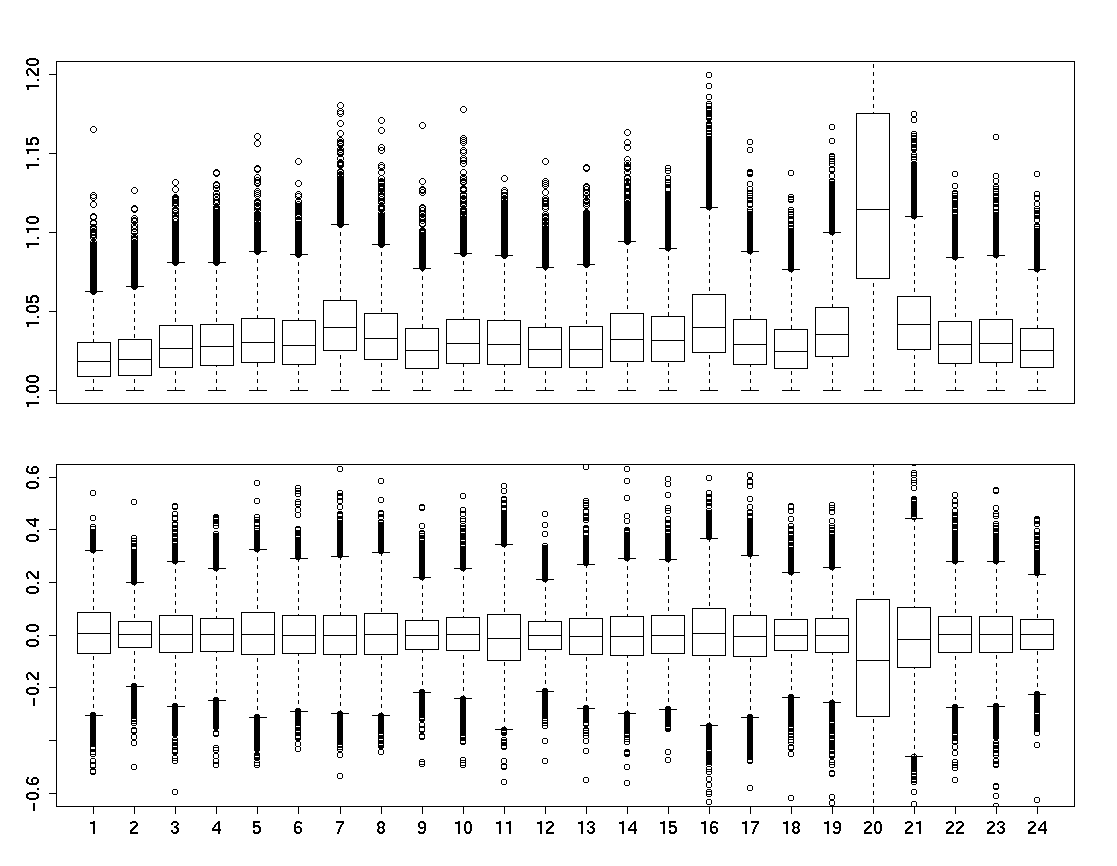}  
\end{center}
\caption
{
{\em
Series of boxplots of NUSE (first row) and RLE (second row) of 24 chips from the Affymetrix HU95 spike-in experiment.  Chip $\#$20 is a clear outlier according to all our quality measures:  Med(RLE), IQR(RLE), Med(NUSE) and IQR(NUSE)\hspace{-2mm}
}
}
\end{figure*}

\newpage

\renewcommand{\thefigure}{A3}
\begin{figure*}[htbp]
\begin{center}
   \includegraphics[width=15cm]{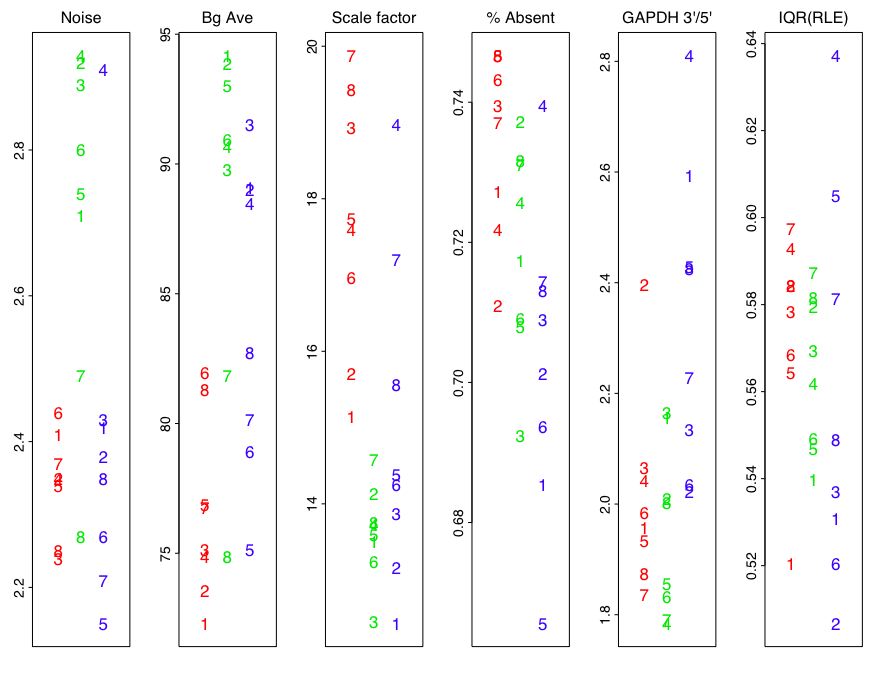} 
\end{center}
\vspace{3mm}
\caption
{{\em
Affymetrix quality scores and IQR(RLE) for the 24 chips from the Affymetrix HU95 spike-in experiment with the 3 technical replicates in different gray levels (left/center/right in each vertical plot).  The outlier chip $\#$20 from Figure A2 is the chip $\#$4 blue in this figure.  It is caught by GAPDH 3'/5'.
By all other Affymetrix quality measures it is placed within the group of lower quality chips, but it does not stand out\hspace{-1.1mm}
}}
\end{figure*}

\renewcommand{\thefigure}{B1}
\begin{figure*}[htbp]
\begin{center}
   \includegraphics[width=15cm]{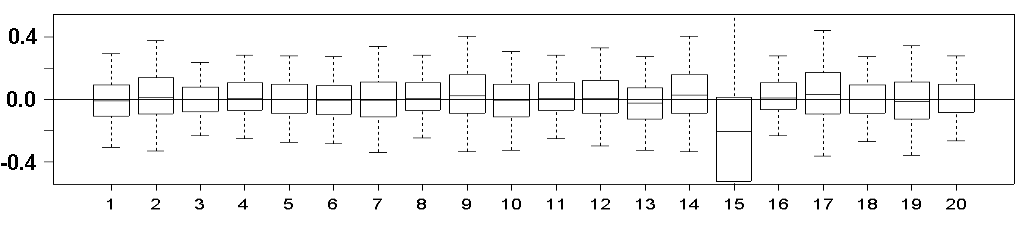}  
   \includegraphics[width=15cm]{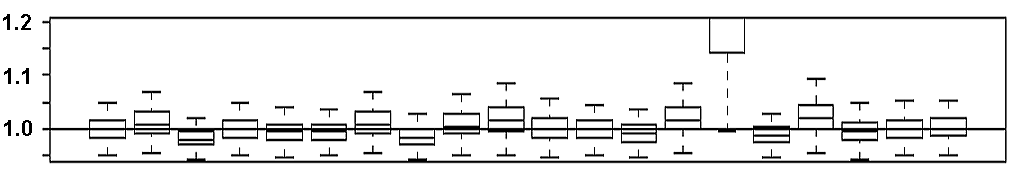} 
\end{center}
\caption
{
{\em
Series of boxplots of RLE (first line) and NUSE (second line) for all 20 MLL 133B chips from {St.~Jude's.}  Chip $\#$15 is an outlier according to all our quality measures:  Med(RLE), IQR(RLE), Med(NUSE) and IQR(NUSE)\hspace{-1.9mm}
}
}
\end{figure*}

\renewcommand{\thesubfigure}{{Chip $\#$}\arabic{subfig}.}

\renewcommand{\thefigure}{B2}
\begin{figure*}[htbp]
  \begin{center}
    \mbox{
      \hspace{-4mm}
      \setcounter{subfig}{10}
      \subfigure[]{ { \includegraphics[width=4.1cm]{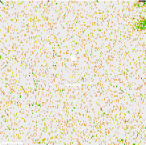} }}
      \hspace{-2mm}
      \setcounter{subfig}{11}
      \subfigure[]{ { \includegraphics[width=4.1cm]{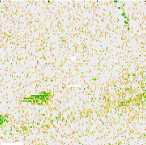} }}
      \hspace{-2mm}
      \setcounter{subfig}{15}
      \subfigure[]{ { \includegraphics[width=4.1cm]{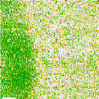} }}
      }
 \end{center}
 \caption
{{\em
Weight images for some of {St.~Jude's} MLL HU133B chips.  Chip $\#$10 and chip $\#$11 show low weights overall and small areas with higher weights.  Chip $\#$15 has elevated weights overall and a region covering almost a third of the total area with extremly high weights\hspace{-1.3mm}
}}
\end{figure*}

\renewcommand{\thefigure}{B3}
\begin{figure*}[htbp]
\begin{center}
   \includegraphics[width=15cm]{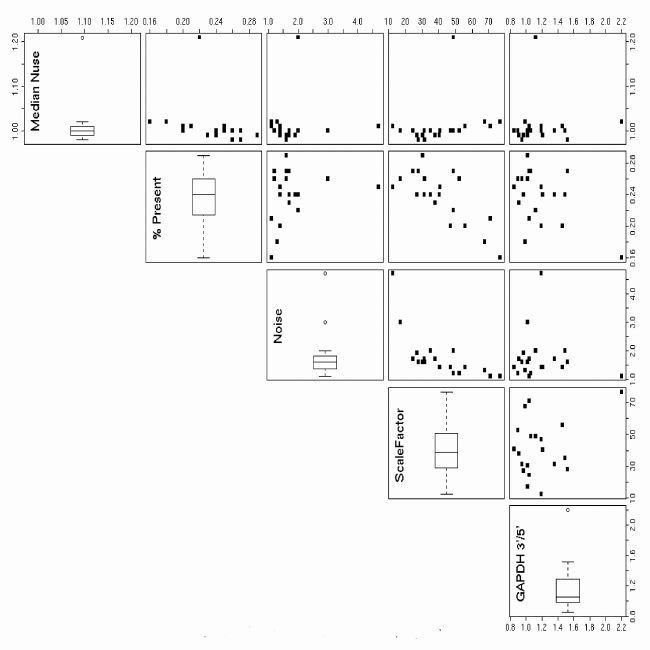} 
\end{center}
\caption
{{\em
Med(NUSE) versus GCOS quality report scores for 20 MLL HU133B chips from St.~Jude's data.  The outlier chip $\#$15 is in the normal range according to all GCOS quality scores\hspace{-1.3mm}
}}
\end{figure*}

\renewcommand{\thefigure}{C1}
\begin{figure*}[htbp]
\begin{center}
      \includegraphics[width=15cm]{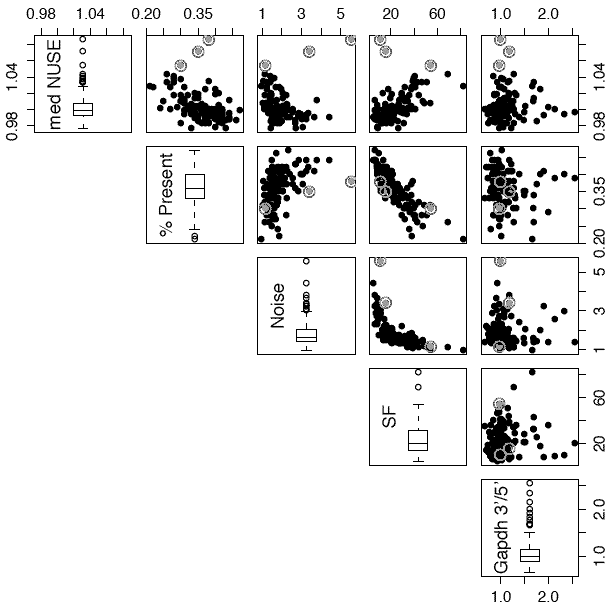} 
\end{center}
\caption
{{\em
Med(NUSE) versus GCOS quality report scores for 129 St.~Jude's HU133A chips. 
Chips with top three Med(NUSE) values are highlighted.
Med(NUSE) shows a linear association with Percent Present and with Scale Factor.  Noise shows some association with Scale Factor, though not quite a linear one.  However, the association between Noise and Med(NUSE) is very weak.  GAPDH 3'/5' 
does not show a linear association with any of the GCOS scores nor with Med(NUSE)\hspace{-1.3mm}
}}
\end{figure*}

\renewcommand{\thesubfigure}{{\em Figure {D}\arabic{subfig}.}}

\renewcommand{\thefigure}{" "}
\begin{figure*}[htbp]
  \begin{center}
    \mbox{
      \hspace{-10mm}
      \setcounter{subfig}{1}
      \subfigure[
Boxplots of qualilty measures for {St.~Jude's} Hyperdip  HU133A chips.  Perfect match distributions are heterogeneous. $\#$12, $\#$8 and $\#$14 have elevated IQR(RLE).  $\#$12 is a clear outlier in NUSE.  The bulk of the chips are of good quality according to both NUSE and RLE.
]
      { { \includegraphics[width=70mm]{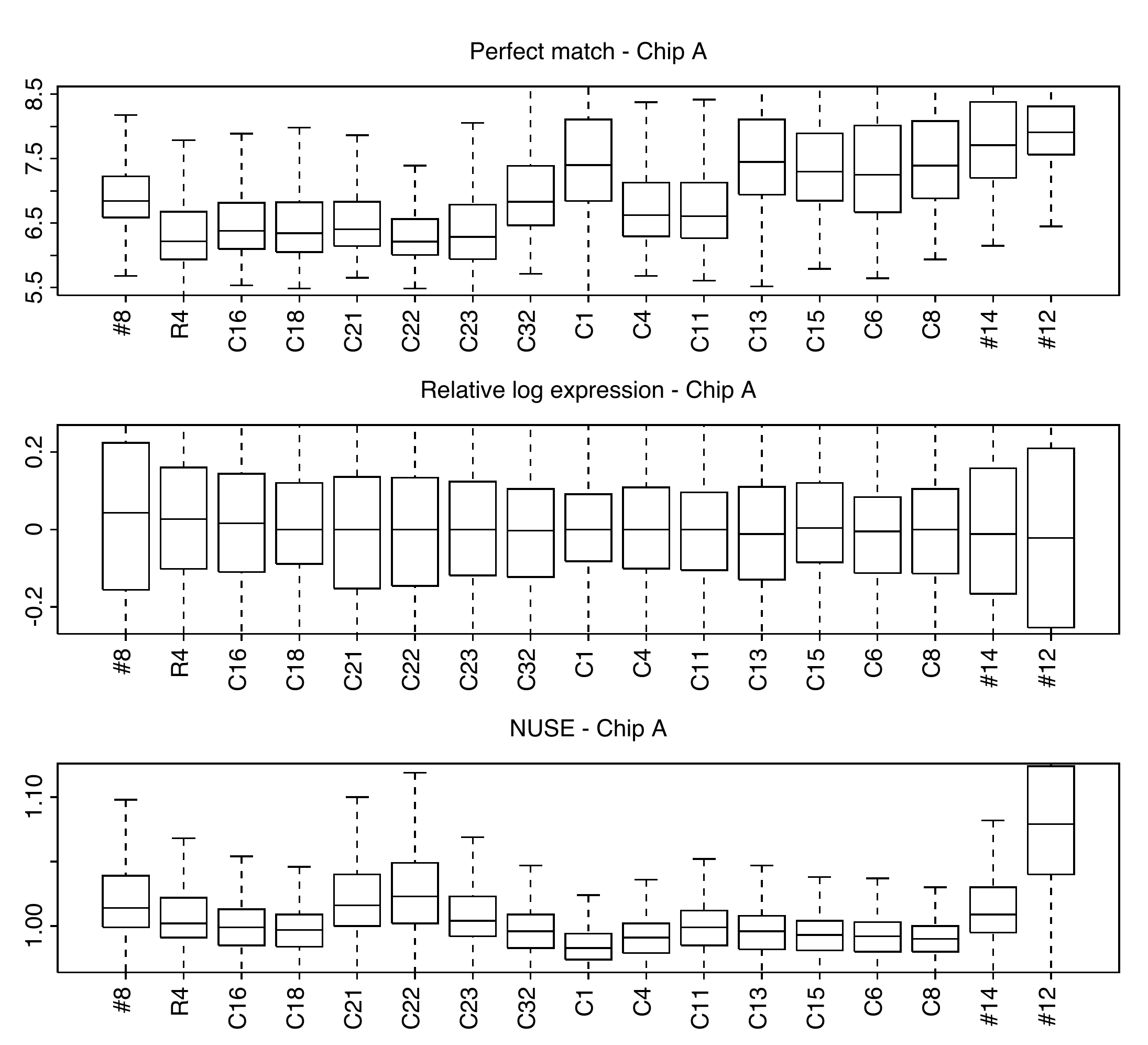} }}
      \hspace{-4mm}
      \setcounter{subfig}{2}
      \subfigure[
Boxplots of qualilty measures for {St.~Jude's} Hyperdip  HU133B chips.  Perfect match distributions are heterogeneous.  $\#$12 has elevated IQR(RLE).  $\#$12 is a clear outlier in NUSE.  The bulk of the chips are of good quality according to both NUSE and RLE.
]{ { \includegraphics[width=70mm]{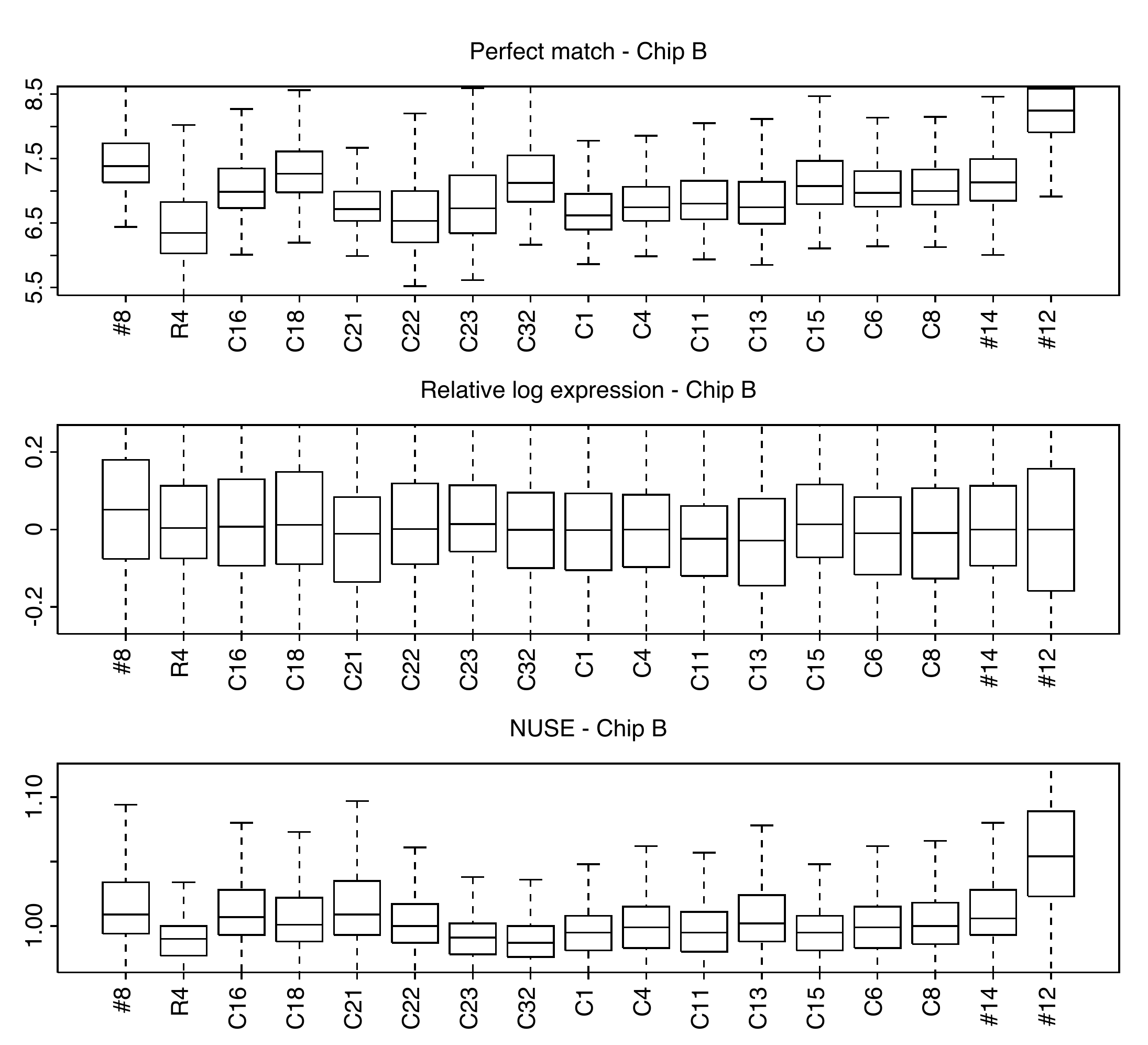} }}
        }
 \end{center}
 \end{figure*}

\renewcommand{\thefigure}{E1}
\begin{figure*}[htbp]
\begin{center}
   \includegraphics[width=6cm]{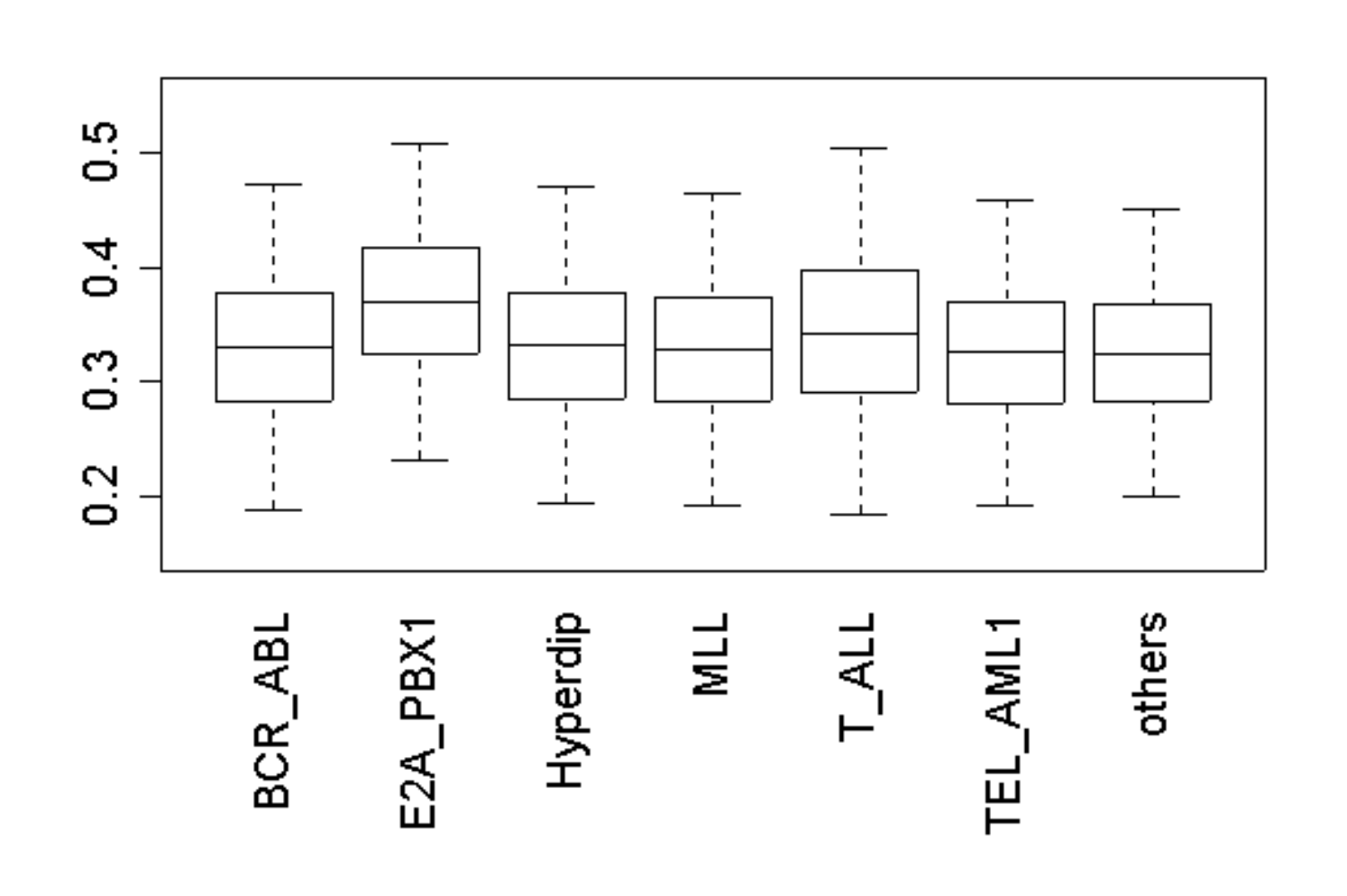} 
\end{center}
\caption
{{\em
Boxplots of the estimated residual scales for diagnostic subgroups in the St.~Jude's dataset.  They show noticable quality differences\hspace{-1.3mm}
}}
\end{figure*}

\renewcommand{\thefigure}{F1}
\begin{figure*}[htbp]
\begin{center}
   \includegraphics[width=15cm]{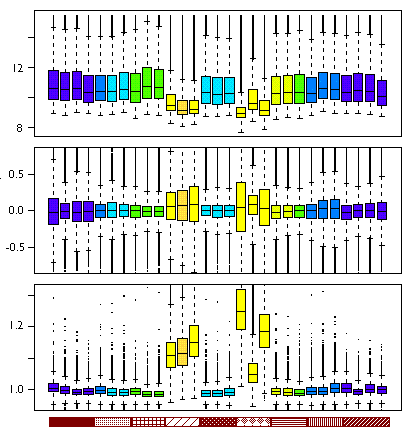}   
  \end{center}
\vspace{-6mm}  
\caption
{{\em
Series of boxplots of log-scaled PM intensities (first row), RLE (second row) and NUSE (third row) for a comparison of nine fruit fly mutants with 3 to 4 technical replicates each.  The patterns below the plot indicate mutants and the colors of the boxes indicate hybridization dates.  Med(RLE), IQR(RLE), Med(NUSE) and IQR(NUSE) all indicate substantially lower quality on the day colored indigo\hspace{-1.3mm}
}}
\end{figure*}

\renewcommand{\thefigure}{F2}
\begin{figure*}[htbp]
\begin{center}
   \includegraphics[width=3.1cm]{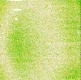} 
\end{center}
\vspace{-2mm}  
\caption
{
{\rm
{``Waves". }
 Weight image for a typical chip in the fruit fly mutant dataset.
 \hspace{-1.3mm} \hfill\newline \qquad 
}
}
\end{figure*}

\renewcommand{\thefigure}{G1}
\begin{figure*}[htbp]
\begin{center}
   \includegraphics[width=14.4cm]{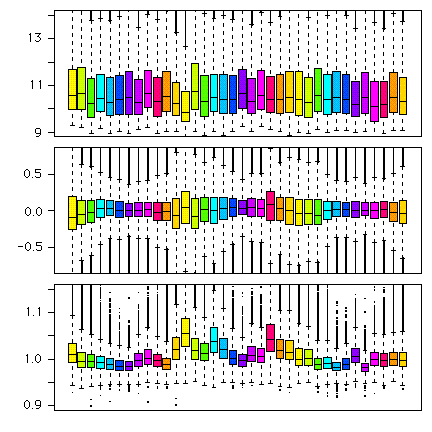} 
\end{center}
\caption
{{\em  
Series of boxplots of log-scaled PM intensities (first row), RLE (second row) and NUSE (third row) in the fruit fly development data.  Boxes 1-12 represent 12 developmental stages (visualized by different colors), boxes 13-24 and boxes 25-36 represent replicates of the whole series.  Replicate series B (boxes 13-24) was hybridized on a different day than the other two series.   Both RLE and NUSE indicate lower quality for series B.  The first couple of chips and the last couple of chips of each series are different from the rest in terms of Med(RLE) deviating from zero (in varying direction),  IQR(RLE) being larger,  Med(NUSE) being elevated.  This is probably mainly driven by the increased high biological variability in these stages\hspace{-1.3mm}
}}
\end{figure*}

\renewcommand{\thefigure}{G2}
\begin{figure*}[htbp]
\begin{center}
   \includegraphics[width=15.2cm]{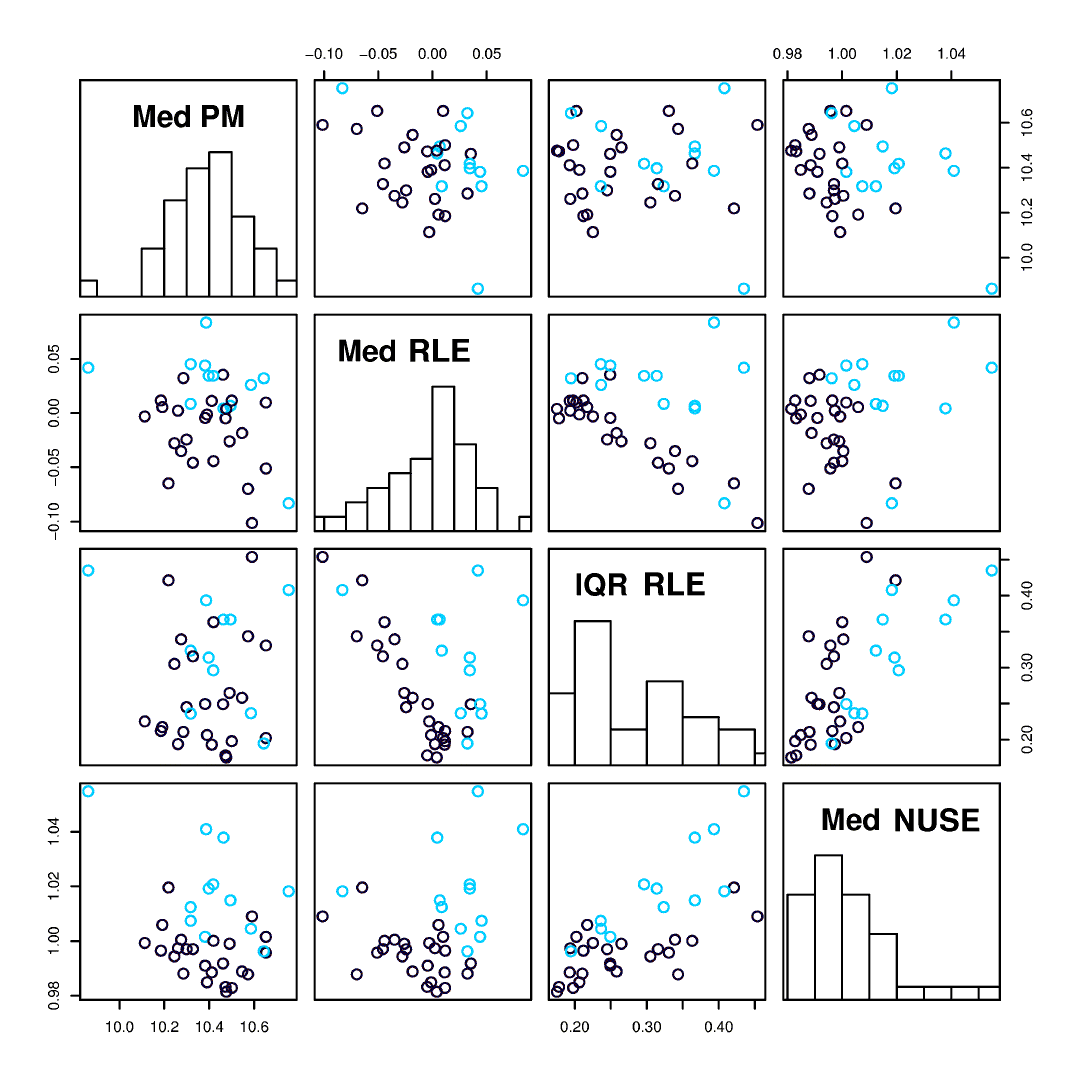} 
\end{center}
\caption
{{\em
Paired quality measures for fruit fly development with chips colored depending on the day of the hybridization.  Med(PM) has no linear association with any of the other measures.  
IQR(RLE) and Med(NUSE) show a linear association on the day colored in light blue and a weak linear association on the day colored in black\hspace{-1.3mm}
}}
\end{figure*}

\renewcommand{\thefigure}{H1}
\begin{figure*}[htbp]
\begin{center}
     \hspace{-15mm} \includegraphics[width=15cm]{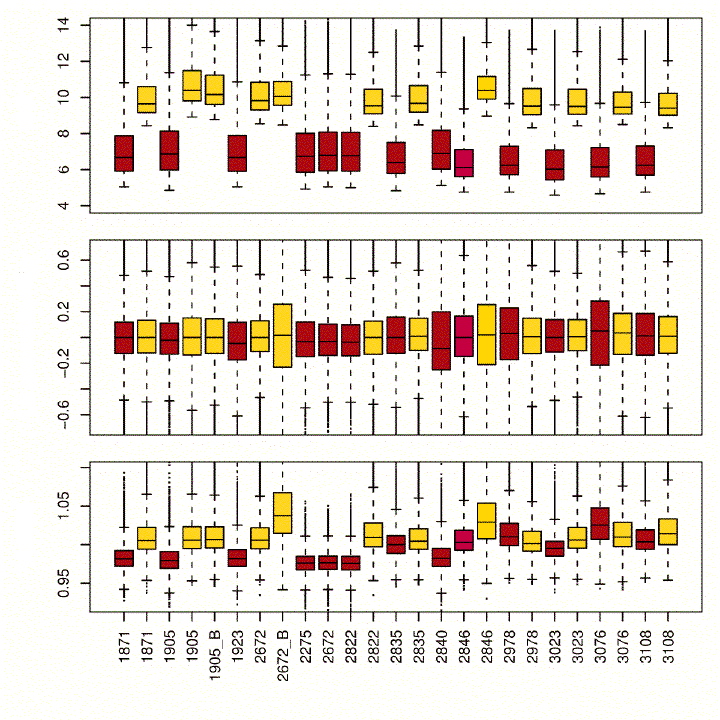} 
  \end{center}

\caption
{{\em
Series of boxplots of log-scaled PM intensities (first row), RLE (second row) and NUSE (third row) for Pritzker gender study brain samples hybridized in two labs (some replicates missing).  Color indicates lab site (dark for lab M, light for lab I).  The PM intensity distributions are all located around 6 for lab M, and around 10 for lab I.  These systematic lab site differences are reflected by IQR(RLE), Med(NUSE) and RLE(NUSE), which consistently show substantially lower quality for lab hybridizations than for lab M hybridizations\hspace{-1.3mm}
}}
\end{figure*}

\renewcommand{\thefigure}{I1}
\begin{figure*}[htbp]
\begin{center}
   \includegraphics[width=15cm]{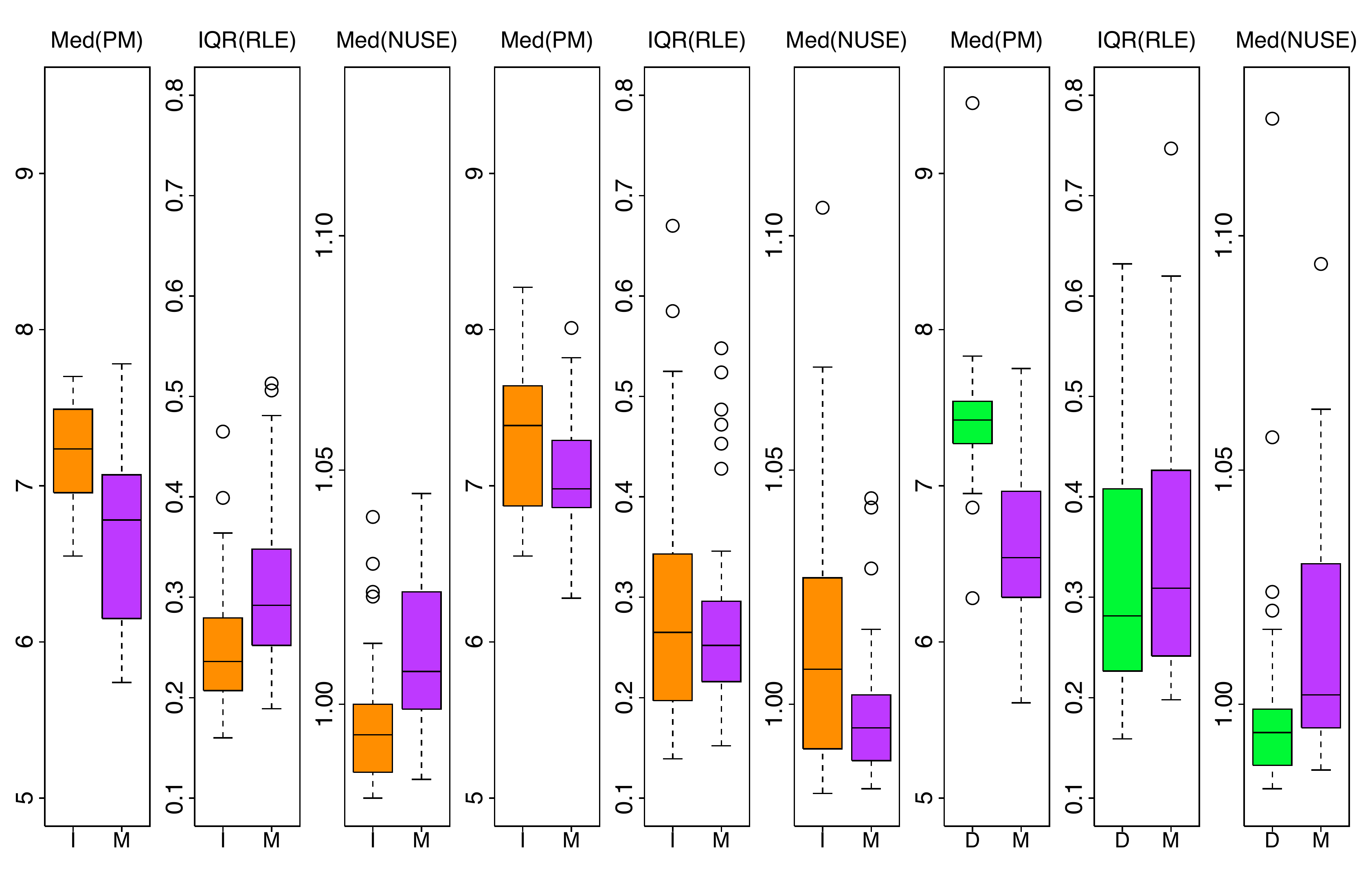} 
\end{center}
\caption
{{\em
Three brain regions from Pritzker mood disorder dataset, each sample hybridized in two labs (lab site indicated by the letters I, M and D, as well as by colors of the boxes).  The boxplots sketch the distribution of quality measure summaries across entire sets of about 40 chips each.  There are still differences between the Med(PM) of the two different labs,  but they are relatively small.  In cerebellum, lab M still produced slightly higher quality hybridizations than lab I, while the ranking is reversed and more pronounced in anterior cingulate cortex.  In dorsolateral prefrontal cortex, lab M shows slightly lower quality than the replicate site D\hspace{-1.3mm}
}}
\end{figure*}

\renewcommand{\thesubfigure}{{J}\arabic{subfig}.}

\renewcommand{\thefigure}{J1-8}
\begin{figure*}[htbp]
  \begin{center}
    \mbox{
      \hspace{-6mm}
      \setcounter{subfig}{1}
      \subfigure["Bubbles"]{ { \includegraphics[width=3.6cm]{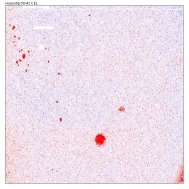} }}\hspace{-5mm}
      \setcounter{subfig}{2}
      \subfigure["Circle and Stick"]{ { \includegraphics[width=3.6cm]{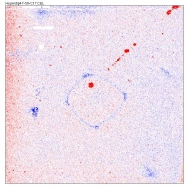} }}\hspace{-5mm}
      \setcounter{subfig}{3}
      \subfigure["Sunset"]{ { \includegraphics[width=3.6cm]{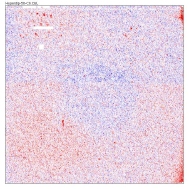} }}\hspace{-5mm}
      \setcounter{subfig}{4}
      \subfigure["Pond"]{ { \includegraphics[width=3.6cm]{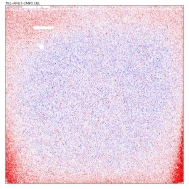} }}
    }
         \mbox{
            \hspace{-6mm}
       \setcounter{subfig}{5}
     \subfigure["Letter S"]{ { \includegraphics[width=3.6cm]{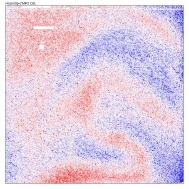} }}\hspace{-5mm}
      \setcounter{subfig}{6}
     \subfigure["Compartments"]{ { \includegraphics[width=3.6cm]{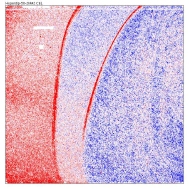} }}\hspace{-5mm}
      \setcounter{subfig}{7}
         \subfigure["Triangle"]{ { \includegraphics[width=3.6cm]{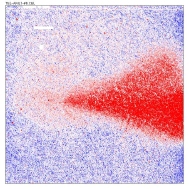} }}\hspace{-5mm}
          \setcounter{subfig}{8}
      \subfigure["Fingerprint"]{ { \includegraphics[width=3.6cm]{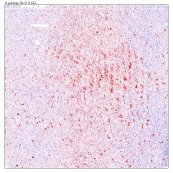} }}
    }
    \caption{{\em Quality landscapes of some selected early St.~Jude's chips\hspace{-0mm}}}
  \end{center}
\end{figure*}

\end{document}